\documentclass[showpacs,aps,prd,floatfix,amsmath,amssymb,nofootinbib]{revtex4}
\usepackage{cleveref}
\usepackage{multirow}
\usepackage{amsmath}
\usepackage{amsfonts}
\usepackage{amssymb}
\usepackage{subfigure}
\usepackage[utf8]{inputenc}
\usepackage{graphicx}
\usepackage{color}
\usepackage{dcolumn}
\usepackage{bm}
\usepackage{float}
\usepackage{footnote} 
\catcode`\@=11
\def\lsim{\mathrel{\mathpalette\@versim<}}
\def\gsim{\mathrel{\mathpalette\@versim>}}
\def\@versim#1#2{\vcenter{\offinterlineskip
\ialign{$\m@th#1\hfil##\hfil$\crcr#2\crcr\sim\crcr } }}
\catcode`\@=12
\def\be{\begin{equation}}
\def\ee{\end{equation}}
\def\bea{\begin{eqnarray}}
\def\eea{\end{eqnarray}}

\newcommand{\nn}{\nonumber}

\newcommand{\yd}{{{Y^d_1}}}

\newcommand{\yyuu}{{{Y^u_2}}}
\newcommand{\hggs}{\Phi_1}
\newcommand{\hggss}{\Phi_2}

\newcommand{\hgtt}{\tilde{\Phi}_2}
\newcommand{\ga}{{\gamma}}

\begin{document}
\title{The Higgs Profile in the Standard Model and Beyond} 
\author{J. Lorenzo D\'iaz-Cruz}
\affiliation{
   CIFFU \\
   Facultad de Ciencias F\'{\i}sico-Matem\'aticas, Benem\'erita Universidad Aut\'onoma de Puebla, Puebla, M\'exico.
}
\begin{abstract}
We present a review of Higgs physics in the SM and beyond, including  the tests of the Higgs boson properties that have been 
performed at LHC and have permitted  to delineate its profile. After presenting the essential features of the Brout-Englert-Higgs (BEH) mechanism, and its implementation in the SM,
we discuss how the Higgs mass limits developed over the years. These constraints in turn helped to classify the Higgs phenomenology
(decays and production mechanisms), which provided the right direction to search for the Higgs
particle, an enterprise that culminated with its discovery at LHC. So far, the constraints on the couplings of the Higgs particle, 
point towards a SM interpretation. However, the SM has open ends that suggest the need to look for extensions of the model. 
 We discuss in general the connection of the Higgs sector with some new physics (e.g. supersymmetry, flavor  and Dark matter),
 with special focus on a more flavored Higgs sector. Thus is realized in the most general 2HDM, and its textured version, 
 which we study in general, and for its various limits, which contain distinctive flavor-violating signals that  could be searched at current and future colliders.
\end{abstract}
\maketitle

\section{Introduction}

The discovery of the Higgs boson, announced on July 4th by the CERN LHC collaborations 
\cite{higgs-atlas:2012gk,higgs-cms:2012gu},  marked the completion of our current theory of the fundamental particles and their
interactions, the so-called Standard Model (SM) \cite{Glashow:1961tr,Weinberg:1967tq,Salam:1968rm}. This event 
can be considered one of the greatest accomplishments of the High Energy Physics community and knowing how it was made possible, as well
as the implications, should become part of the culture of particle physics. This is precisely the purpose of this review paper.

The search for an understanding of the structure of matter and its interactions, has been a moto of the physical sciences, which could be 
traced back to the early days of modern atomic theory. After the establishment of quantum mechanics and its application
to atomic and nuclear systems, relativist quantum fields made its appearance and allowed to formulate a consistent quantum  theory for electrons and photons, quantum electrodynamics (QED), which successfully predicted antimatter (positrons for the electron). Afterwards, particle physics passed through great times in the 50-60's, when plenty of data was collected 
by the experimental collaborations from high-energy labs around the world. This branch of physics advanced by the detection of a rich spectrum of new strongly interacting resonances, the hadrons. However, it was not known whether quantum field theory (QFT) would survive or not, 
as the correct description of particles and interactions.  More general approaches, such as dual models, saw the light just to be shown years later to be a different guise  for same good old QFT. 

Progress was also made on the description of Weak interactions, which move from the Fermi 4-fermion effective theory
of beta decay, to the Intermediate Vector Boson theory. Promising quantum unified theories were formulated, which included non-abelian symmetries that contained self-interacting gauge bosons.  These theories seemed to work only for massless gauge fields, as the photon of QED, but in the case of weak interactions it was needed to consider massive gauge bosons (such as the charged $W^{\pm}$ that was supposed to mediate the weak interactions). 

Two lines of work were undertaken, one of them attempted to find some consistent way to make the Yang Mills fields massive,
independently of a realistic model.
Following the ideas of Nambu  on spontaneous symmetry breaking, a solution to this problem appeared 
with the initial formulation of the Brout-Englert-Higgs (BEH) mechanism  \cite{Higgs:1964pj,Englert:1964et,Guralnik:1964eu}, 
which was shown to be a viable mechanism to generate the masses of the elementary particles (gauge bosons and chiral fermions). This mechanism was incorporated within the unified Electroweak (EW) gauge model \cite{Glashow:1961tr,Weinberg:1967tq,Salam:1968rm}.
From the model building perspective, the electroweak SM was completed after the Glashow-Iliopoulos-Maiani mechanism was proposed \cite{Glashow:1970gm}, 
which received further confirmation with the detection of the charm quark. The strong interactions were described by a gauge theory too, quantun chromodynamics (QCD), where quarks bind into hadrons through interactions mediated by gluons.

The other line involved a detailed study of the renormalization problem associated with 
having massive gauge bosons, with M. Veltman acting as the driving force behind this project \cite{Veltman:1968ki}.
Nowadays, we know that the SM is a renormalizable Quantum Field Theory (QFT), that is quite successfull in describing 
the interactions of the fundamental particles. 
The detection of the W and Z in the 80's at CERN, and jet events initiated by gluons at DESY, brought extra confirmation of the SM. 

The Higgs boson was recognized as a testable sector the SM in the mid 70's and early 80's, when the first 
papers discussed the methods to  search for direct effects of the Higgs boson  in particle accelerators 
\cite{Ellis:1975ap, Georgi:1977gs, Glashow:1978ab}, and this was complemented by the study of indirect Higgs effects 
(Radiative corrections)  on the properties of the SM particles and its related parameters \cite{Veltman:1977fy}.
{\footnote{For reviews of early papers on Higgs hunting, see Refs. \cite{Gunion:1989we, Carena:2002es, Djouadi:2005gi}.}}

Given the success of the SM, it became credible that all interactions could be unified in a single gauge group. Construction of experimental facilities to search for proton decay were started in the early 80's, but since no proton decay was observed, the unification paradigm loosed some momentum. At that time a concrete Supersymmetric (SUSY) extension of the SM was proposed too, which was shown to ameliorated the hierarchy/naturalness problem of the Higgs mass. 

This is more or less the time when the author entered into graduate school, at the University of Michigan, and continued studying the Higgs particle.  The community was engaged in a debate about the best options for constructing of next colliders, and 
there were discussions about the best strategy to produce and detect the Higgs particle, although many efforts were devoted to build models that did not require an elementary scalar, the technicolor adventure.  Given the success of the hadron machine that  made possible to detect the W and Z bosons, it seemed that the best strategy for a high energy collider, was the proton-proton or proton-antiproton  choices.  The  USA made plans to study the  Higgs particle, and other forms of new physics, 
at a proton-proton collider, the SSC, that was designed to work with c.m. energy of  40 TeV. 
Tevatron (a proton-antiproton collider) was also approved, with c.m. energy of 1.8 TeV, which was thought to be good enough
to produce and detect the top quark.
But this was the mid 80's, and CERN had planned to build a circular electron-positron collider
that would work first at the Z pole (LEPI) and then above the WW threshold (LEPII). 

The reason why such high energy was considered for SSC, was in part due to the lack of knowledge of the Higgs mass.
At U of Michigan, a series of talks were planned for the fall of 87 to discuss the search for the Higgs boson. 
First talk was by Tiny Veltman, who argued that the best way to start getting information on the Higgs mass was 
from a detailed analysis of radiative corrections,
which years later proved to be very useful to constrain the Higgs mass. The second talk was by Gordy Kane, who presented the 
different techniques that should be used to search for a Higgs particle at a hadron collider. In those
days the Higgs mass ($m_h$) was classified into several ranges: light ($m_h \leq m_Z$), intermediate ($ m_Z \leq m_h \leq 2m_t$) 
and heavy $2 m_t \leq m_h \leq 600$ GeV). Considering these mass ranges made sense at that time, because it was thought that 
the top mass  should be close to the  value of 45 GeV, as was claimed early on by C. Rubbia at CERN.
As the Tevatron entered into the game, the lower limits on  the top mass started to increase, 
and together with results from B-physics, it was suspected that $m_t$ was above 90 GeV. Until it was  detected at
Tevatron in the mid 90's with mass $m_t \simeq 175$ GeV, and then the intermediate Higgs mass range was redefined
as $ m_Z \leq m_h \leq 2m_Z$.

A light Higgs could have been detected using the associated production of the Higgs with a gauge boson, with the Higgs decaying
into bb pairs. In fact, such range was the task of the LEP collider at CERN. The most difficult task, as Kane argued, 
was the above mentioned   intermediate mass range.  After we learned that top was heavier  than $m_Z$, the intermediate mass range was redefined, and its associated difficulties disappeared from Higgs hunting considerations, but the techniques that were devised by Kane et al, proved to be very useful in the ultimate search at LHC that provided first hints of the Higgs particle in 2012.  
Heavy higgs could had been detected in the golden mode $h\to ZZ$, until the Higgs mass
was so large that a perturbative treatment would no longer be reliable. Such range was termed the obese mass region, starting from about 600 GeV.  The third talk of the series was presented by an experimentalist, R. Thun, who discussed the issue of the signal vs backgrounds, with some estimated characteristics for semi-realistic detectors, and he showed that both theoretical works would be very difficult to realize at planned colliders.

The 90's witnessed the entering of operation of LEP II, with cm energy of 200 GeV. The
search for Higgs at LEP used the reaction $e^+e^- \to h+Z$, and resulted in successive bounds on Higgs mass, which 
eventually reached the value $m_h \geq 111$ GeV
at the end of LEP life-time. The tail of a Higgs signal was believed to had been observed at the final stages of LEP,
which ignited some pressure to keep LEP running, but the director say no, which was in fact a good choice, and the tunnel was cleaned
to start the installation of LHC magnets and the construction of the giant detectors ATLAS and CMS. These detectors 
were designed to catch a Higgs boson in the narrow mass range left by the analysis of electroweak precision tests, namely
$115 \leq m_h \leq 130$ GeV, which could be probed with the modes $h\to \gamma \gamma$ and $h\to ZZ$ that were studied
early on by Kane et al 
{\footnote[2]{So, in the end  both Kane and Veltman were right about their strategies for
Higgs physics}}.

LHC started taking data on 2011, but nothing really big came out of those early runs apart from some fake signals and rumors. 
The big news have to wait until mid 2012, when the LHC  announced the discovery of a Higgs-like particle with $m_h=125-126$ GeV at the LHC \cite{higgs-atlas:2012gk,higgs-cms:2012gu}, 
a landmark event that provided a definite test of the mechanism of Electro-Weak symmetry breaking \cite{Gunion:1989we}. 
Thus, after many years of hypothesis and conjectures, it was finally possible to confirm that "what we thought about
the origin of masses and Higgs mechanism,  is real ..." \cite{kane}.

The Higgs  mass  value agrees quite  well with the range preferred by the 
electroweak precision tests (EWPT)  \cite{Erler:2007sc}, which confirms the success of the SM. 
Current measurements of its spin, parity, and couplings,  also seem consistent with 
the SM. The fact that LHC has verified the linear realization of spontaneous symmetry 
breaking (SSB), as included in the Standard Model (SM), could also be taken as an
indication that Nature likes scalars. 

The initial reports from LHC claimed the discovery of a resonance with mass $m=125-126$ GeV,
through its decays into $\gamma\gamma$ and $ZZ^{*}$, which were consistent with having either spin $s=0$ or $s=2$.
Later on, with more data collected from more decay modes, it was concluded that the simplest choice $s=0$ was favored. 
After LHC delivered the Higgs signal, many papers have been devoted to study
the Higgs couplings, and the constraints on deviations from SM   
\cite{Espinosa:2012im, Cheung:2013rva, Ellis:2014dva}. 
Although the initial data showed also some tantalizing hints of deviations from the SM predictions, including first a possible enhanced 
$\gamma\gamma$ rate,  and later on a signal from the LFV Higgs decay modes appeared to had been detected, these signals were
not confirmed with more data. 
So far, we can say that the signal resembles a Higgs scalar, with 
a profile consistent with the SM interpretation.

On the other hand, the LHC has also been searching for signals of Physics
Beyond the SM, which has been conjectured in order to address some of the problems 
left open by the SM, such as hierarchy, flavor, unification, etc \cite{Arkani-Hamed:2016kpz}.
Some of these extensions of the SM, such as SUSY and multi-Higgs models in general \cite{Accomando:2006ga}, 
predict deviations from the SM Higgs couplings, while at the same time contain a
rich Higgs spectrum, whose detection would be a clear signal of new physics.

However, the LHC has provided bounds on the new physics scale ($\Lambda$), 
that are already entering into the multi-TeV range, and this is  casting some doubts about
the theoretical motivations for new physics scenarios with a mass scale 
of order TeV. This is particularly disturbing for  
the concept of naturalness, and its supersymmetric implementation, since the bounds 
on the mass of superpartners are passing the TeV limit too.
 However, some of the motivations for new physics are  so deep, that it seems reasonable
to wait for the next LHC runs, with higher energy and luminosity,
in order to have stronger limits, both on the search for new particles, such as heavier
Higgs bosons, and for precision tests of the SM properties.

This review paper is intended to cover the essential of Higgs physics, starting with a quick review of the SM Higgs, then looking at 
the motivations for some extensions of the  Higgs sector, and focusing then in the most general Two-Higgs doublet model (2HDM).
 We hope our paper complements the excellent reviews on Higgs physics that have appeared recently 
\cite{Altarelli:2013lla,Ellis:2015tba,Haber:2017udj,Dawson:2018dcd,Kane:2018oax,Wells:2018nwj}. 
The organization of our paper goes as follows. We shall present in Section 2, the SM Higgs Lagrangian, including the Higgs potential, 
the gauge and Yukawa interactions, as well as the theoretical constraints on the Higgs mass. We start section 3 with discussion of the Higgs couplings with gauge bosons  and fermions, first within the SM and then presenting
a model-independent approach; this parametrization is proposed to describe  Higgs couplings that include flavor 
and Charge-conjugation Parity (CP) violation. 
Then, we discuss briefly the Higgs phenomenology at the LHC, which allowed to gather information 
and draw the current Higgs boson profile. 
 Although the Higgs properties could be discussed within an effective Lagrangian approach,
it is also important to discuss them within an specific model, where such deviations 
could be interpreted and given a context. 
Thus, in section 4 we discuss the motivation for extending the SM Higgs sector, with a focus on
the multi-Higgs models, including its Supersymmetric
versions and models with Higgs-Flavon mixing; some remarks on the Higgs portal and its dark matter (DM) connection
is presented too.
Section 5 contains a detailed discussion of one of such models where one has a "more flavored  Higgs sector", 
namely the most general two-Higgs doublet model
(the 2HDM of type III and its textured realization),
which includes new sources of flavor and  CP violation. As phenomenological predictions we
discuss the Lepton Flavor Violating (LFV) Higgs decays $H_i \to l_i l_j$ and the top  quark 
Flavor-Changing-Neutral-Currents (FCNC) transitions  $t\to ch$ \cite{Hou:1991un}. 
Concluding remarks are included in section 6, where we end with a brief discussion about possible paths for the future,
including some comments about the options to tests of the Higgs couplings with 
light quarks, which in turn take us to consider the so called "private" Higgs models.

\section {\large  SM Higgs Lagrangian: Gauge and Yukawa couplings}

After many years of theoretical and experimental efforts, we have now a great theory of the elementary constituents
of nature and its fundamental interactions, the Standard Model (SM), which is based on the following 
(For recet texts see: \cite{Lancaster:2014pza,Schwartz:2013pla} ):
\begin{itemize}
 \item The fundamental particles are the quarks and leptons, which appear repeated in three chiral families. Quarks form the hadrons, 
 such as the proton, neutron, pions, etc. Charged leptons, such as electron, muon and tau, are accompanied by the light neutrinos.
 \item The quarks and leptons have interactions that follow from the Gauge principle, i.e. the forces are mediated by vector
 particles associated with gauge symmetries, and there is  one gauge field for each  generator of the Lie algebras associated with the
 symmetries of the system.
 \item The masses of the weak gauge bosons ($W^{\pm}, Z$), and the fermions, arise as a consequence of the interactions of the
 particles with the vacuum, which can live in a broken phase, i.e. the   BEH  mechanism.
\end{itemize}

The  BEH mechanism, was implemented by Weinberg in the model proposed by Glashow, which was based on the 
gauge group $SU(2)_L \times U(1)_Y$. Treating quarks and leptons as the fundamental degrees of freedom, lead to 
the correct formulation of the SM. We shall start by presenting the essential features of the SM and the BEH mechanism.

\subsection{SM gauge and fermion sector}

The weak and electromagnetic interactions are partly unified into the Electroweak theory, which has the gauge symmetry
$SU(2)_L\times U(1)_{Y}$. The Lagrangian for the gauge and fermion sector (leptons) is given by:

\begin{equation}\label{lagrangiano_EW}
{\cal{L}}=\bar{L} \ga^{\mu}D_{\mu}L+\bar{e}_R \ga^{\mu}D'_{\mu}e_{R}-\frac{1}{4}W^{\mu\nu i}W_{\mu\nu}^{i}-\frac{1}{4}B^{\mu\nu}B_{\mu\nu},
\end{equation}

where
\begin{align}\label{derivada_covariante1}
D_{\mu}&=\partial_{\mu}+i\frac{g}{2}\sigma^{i}W_{\mu}^{i}+i\frac{g'}{2}B_{\mu},\\\label{derivada_covariante2}
D'_{\mu}&=\partial_{\mu}+ig' B_{\mu},
\end{align}

Here $L$ denotes the left-handed lepton doublet, while $e_R$ corresponds to the right-handed electron, $\sigma^{i}$ are
 the Pauli matrices and $g$ y $g'$ denote the gauge coupling constants. The tensors $W_{\mu\nu}^{i}$ and  $B_{\mu\nu}$ are given by:
\begin{align}
W_{\mu\nu}^{i}&=\partial_{\mu}W_{\nu}^{i}-\partial_{\nu}W_{\mu}^{i}-g\epsilon^{ijk}W_{\mu}^{j}W_{\nu}^{k},\\
B_{\mu\nu}&=\partial_{\mu}B_{\nu}-\partial_{\nu}B_{\mu}.
\end{align}

In addition we need to include the quarks, with the left-handed components forming a doublet of weak isospin, 
while the right-handed ones transform as singlets. Quarks are also triplets of the 
color interactions, which is described by the gauge symmetry $SU(3)_c$.
Thus, the full  SM gauge symmetry is:  $SU(3)_c\times SU(2)_L\times U(1)_Y$.


\subsection{ SSB and the Higgs}

Insights from the role of the vacuum in condensed matter, helped to identify how a symmetry is realized in QFT, namely:
\begin{itemize}
\item  A symmetry is realized a la Wigner-Weyl, when the vacuum respects such symmetry, i.e. the minimum
of the energy happens for vanishing field variables, as in QED where $<A^{\mu}>=0$. 
\item  A realization a la Nambu-Goldstone occurs when the vacuum does
not respect the symmetry of the Lagrangian, which leads to Goldstone Theorem: When a Global symmetry 
is broken spontaneously (SSB), massless particles associated
with the broken generators appear. 
\end{itemize} 

SSB happens for instance in strong interactions, where the Chiral symmetry is not manifest,
and the pions are identified as pseudo Nambu-Goldstone bosons (NGB)
(pseudo means they are not exactly NGB, due to the small mass of the light quarks ($u,d , s$)).

In the case of the weak interactions, the problem was how to include the mass for the charged W bosons,
the mediators of the weak interactions that manifest in neutron beta decay. The solution 
came with the BEH mechanism \cite{Englert:1964et, Higgs:1964pj, Guralnik:1964eu}, 
which proved that within the context of a local symmetry,  the degrees of freedom associated with the 
Goldstone bosons become the longitudinal modes of the gauge bosons,
which then acquire a mass.

Let us discuss this mechanism for a simple Abelian theory, with Lagrangian:

\begin{equation}\label{campo_escalar}
{\cal{L}}_{\phi}=\frac{1}{2}\partial_{\mu}\phi\partial^{\mu}\phi-V(\phi),\hspace{0.5cm}V(\phi)=\frac{1}{2}\mu^2\phi^2+\frac{1}{4}\lambda\phi^4
\end{equation}
where $\lambda>0$. 

This Lagrangian (\ref{campo_escalar}) is invariant under the parity (P) transformation $\phi\to-\phi$. 
However, when one minimizes the potential, we find that:

\begin{itemize}
\item{(a)} For $\mu^2>0$, the minimum is invariant under P, and it occurs for: 
\begin{equation}
\langle\phi \rangle_0=\langle0|\phi|0\rangle=0.
\end{equation}
\item{(b)} For $\mu^2<0$, the minimum is displaced from the origin, and does not respect P, i.e.
\begin{equation}
\langle \phi	\rangle_0=\pm\sqrt{\frac{-\mu^2}{2\lambda}}\equiv\pm\frac{v}{\sqrt{2}}.
\end{equation}
\end{itemize}

Now, there is a degeneracy between  $v$ y $-v$. Then, one can study the fluctuations
around the new minimum, which are interpreted as the particles, i.e.
$\xi(x)\equiv\phi(x)-\langle\phi\rangle_0=\phi(x)-v$.

Then, the Lagrangian for the field $\xi$ (the fluctuation) becomes:

\begin{equation}
{\cal{L}}_{\xi}=\frac{1}{2}\partial_{\mu}\xi\partial^{\mu}\xi-\lambda v^2\xi^2-\lambda v \xi^3-\frac{1}{4}\lambda\xi^4.
\end{equation}

Thanks to SSB, this Lagrangian contains now a scalar field $\xi$ with mass $m_{\xi}= \sqrt{2\lambda}v = \sqrt{-2\mu^2}$. 
 When one uses SSB within the context of a gauge theory, it is possible to generate masses for the gauge bosons.

Within the SM the gauge symmetry for the electroweak interactions is $SU(2)\times U(1)_Y$, with gauge bosons $W^{\pm} _\mu, W^3_\mu$ and $B_\mu$. 
The minimal Higgs sector includes one Higgs doublet, which can be written as follows:
\begin{equation}\label{Higgs_def}
\Phi=\frac{1}{\sqrt{2}}\begin{bmatrix}
    \phi_1  -i\phi_2\\
    \phi_3  -i\phi_4
\end{bmatrix}.
\end{equation}

The mass terms for the $W,Z$ is obtained from the  the Higgs Lagrangian, which includes the Higgs kinetic term, its  gauge interactions,
and the Higgs potential, i.e.
\begin{equation}\label{lagrangiano_BEH}
{\cal{L}}_{H}=|D_{\mu}\Phi|^2-V(\Phi),
\end{equation}

The potential $V(\Phi)$ is written as:
\begin{eqnarray}
V(\Phi)&=& \mu^2 \Phi^{\dagger} \Phi + \lambda \left( \Phi^{\dagger} \Phi \right)^2 \nonumber \\
       &=&\frac{1}{2}\mu^2\left(\sum_{j=1}^{4}\phi_i^2\right)+\frac{1}{4}\lambda\left(\sum_{j=1}^{4}\phi_i^2\right)^2.
\end{eqnarray}

For $\mu^2 < 0$, the gauge symmetry $SU(2)_L\times U(1)_Y$ is broken to $U(1)_{em}$. 
This is illustrated in figure 1, where we can see the different options for the vev (red points), we also display the 3-dimensional image 
 of the potential, i.e. the mexican hat potential.
In this case, part of the scalar degrees  of freedom  from the doublet, the so-called pseudo Goldstone bosons (pGB), 
become the longitudinal modes of the $W^\pm$ and $Z$, which then become massive. The charged pGB is identified as:
$G^{\pm}= \frac{1}{\sqrt{2}} ( \phi_1  \pm i\phi_2) $, while the neutral pGB correspond to: $G^0= \phi_4$. 
The remaining  degree of freedom ($\phi_3$) includes a physical d.o.f., which by itself "forms an incomplete scalar multiplet", 
as Higgs pointed out in his classic paper \cite{Higgs:1964pj}.

\begin{figure}[t]
 \begin{center}
\includegraphics[width=7.7cm, height=7.3cm]{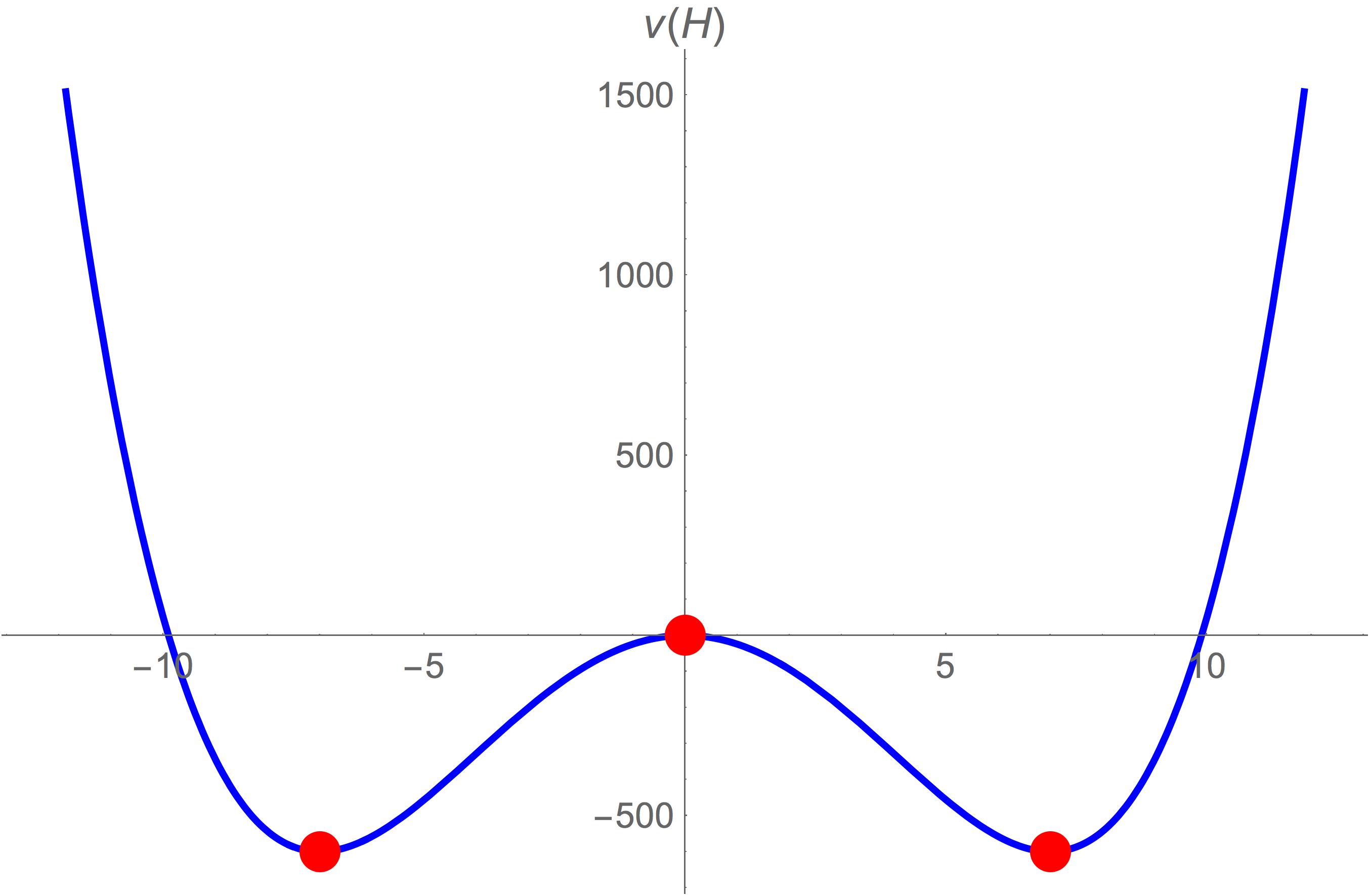}
\includegraphics[width=7.7cm, height=6.3cm]{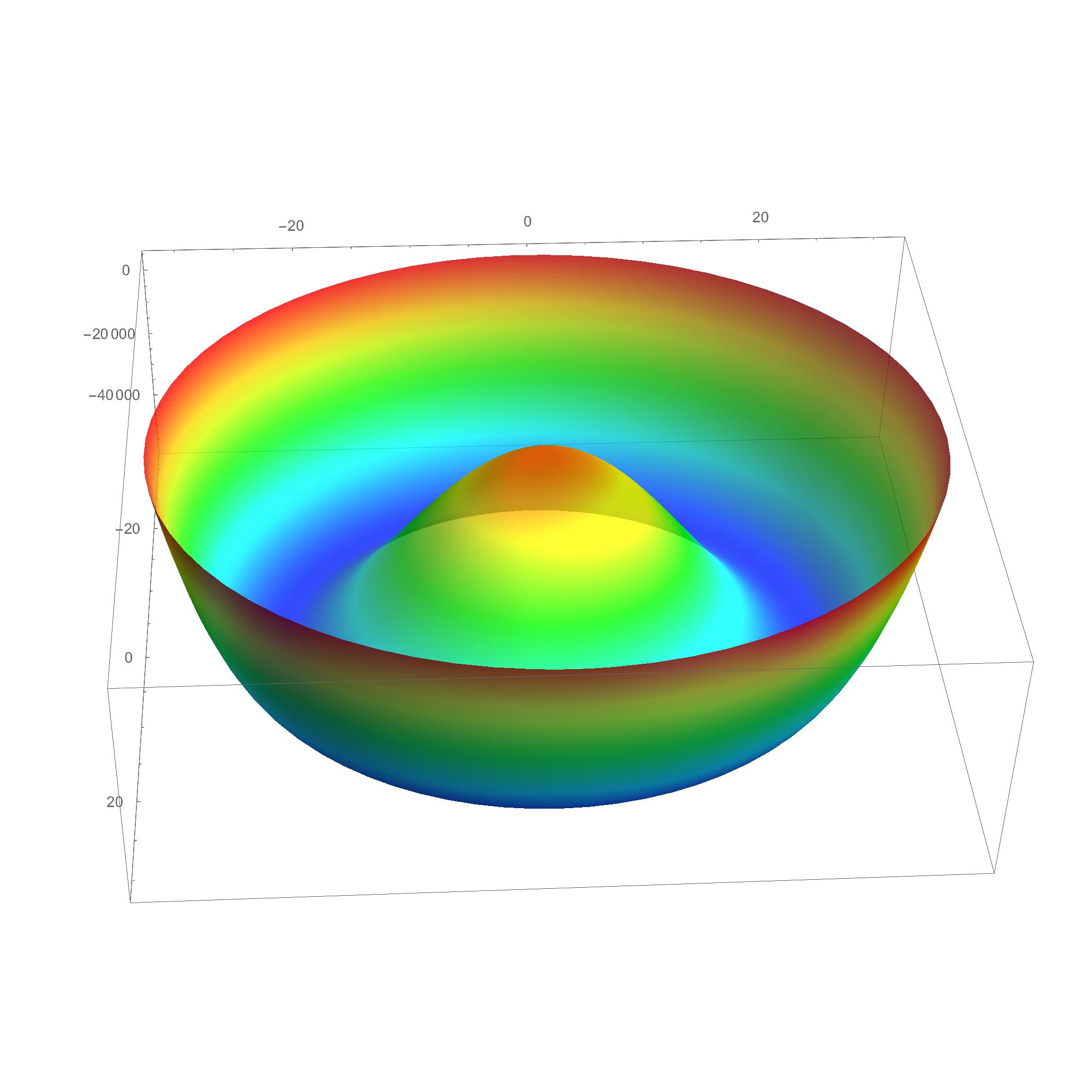}
\caption{The Higgs potential when SSB occurs (left), and the 3-dimensional image, i.e. the mexican hat potential (right).}
\label{Fig:HiggsPot}
\end{center}
\end{figure}

However, the gauge symmetry leaves us some freedom to remove non-physical degrees of freedom. This is 
done in the so called Unitary gauge, where one can take:
$\phi_1=\phi_2=\phi_4=0$ y $\phi_3=v+h$; and $h$ corresponds to the excitations from the vev.
Thus, the Higgs doublet $\Phi$ can be written in the unitary gauge as:
\begin{equation}\label{Higgs_def2}
\phi=\frac{1}{\sqrt{2}}\begin{bmatrix}
    0\\
   v+h
\end{bmatrix},
\end{equation}

Then, the potential $V(\phi)$ becomes:
\begin{equation}
V=\frac{1}{4}\lambda v^4+\lambda v^2 h^2+\lambda v h^3+\frac{1}{4}\lambda h^4.
\end{equation}

The second term in the potential (quadratic in h) indicates that the Higgs ($h$) mass is $m_h^2=2v^2\lambda$. 
The cubic and quartic terms ($h^3, h^4$) describe the 3- and 4-point Higgs self-couplings. 

On the other hand, the interaction of the Higgs with the gauge bosons, is contained in the 
kinetic term, which can be written as:

\begin{equation}\label{kinetic_term}
\bigg|\left(i\frac{g}{2}\sigma^iW_{\mu}^i+i\frac{g'}{2}B_{\mu}\right)\Phi\bigg|^2=\frac{(v+h)^2}{8}\left(g^2(W^{1}_{\mu})^2+g^2(W^{2}_{\mu})^2+(-gW^3_{\mu}+g'B_{\mu})^2\right).
\end{equation}

We define new vector fields:

\begin{align}\label{gauge_boson1}
Z_{\mu}^0&=\frac{1}{\sqrt{g^2+g'^2}}(gW^{3}_{\mu}-g'B_{\mu}),\\\label{gauge_boson2}
W_{\mu}^{\pm}&=\frac{1}{\sqrt{2}}(W_{\mu}^1\pm i W_{\mu}^2),\\\label{gauge_boson3}
A_{\mu}^{\pm}&=\frac{1}{\sqrt{g^2+g'^2}}(g'W^{3}_{\mu}+gB_{\mu}),
\end{align}

Thus, the Lagrangian (\ref{kinetic_term}) contains mass terms for these fields, i.e. 
$M_W^2W_{\mu}^{+}W^{-\mu}+\frac{1}{2} M_Z^2Z_{\mu}Z^{\mu}$.
The corresponding masses given by
$M_{W}=\frac{vg}{2} ,\hspace{0.4cm} M_{Z}=\frac{v}{2}\sqrt{g^2+g'^2}$,
while the photon remains massless, i.e. $M_A=0$.

We can also express the electric charge ($e$) in terms of the gauge couplings  $g$ y $g'$, as
follows: $e=\frac{g g'}{\sqrt{g^2+g'^2}}= g \sin\theta_W$, where $\theta_W$ denotes the weak mixing angle. 
Within the SM, there holds a relationship between the gauge boson masses, which turns out to be of crucial  importance, 
i.e.  $M_W=M_Z\cos\theta_W$, with $\cos\theta_{W}=\frac{g}{\sqrt{g^2+g'^2}}$ and $\sin\theta_W=\frac{g'}{\sqrt{g^2+g'^2}}$. 
Finally, one can also expres the Fermi constant as:
$G_F=\sqrt{2}\frac{g^2}{8M_{W}^2}=\frac{\sqrt{2}}{2v^2}$.

The numerical value of the fermi constant is: $G_F=1.16637\times10^{-5}GeV^{-2}$ \cite{PDG}, which implies:
$v\sim246$ GeV, this value is known as the Electroweak scale. 
One can check that the interactions of the type
$hVV$ and $hhVV$ are contained in this Lagrangian, namely:
\begin{equation}
 {\cal{L}}_{hVV}= g M_W h W_{\mu}^{+}W^{-\mu}+\frac{gM_Z}{2 c_W}  h Z_{\mu}Z^{\mu},
\end{equation}

In the SM, the fermions are chiral fields, i.e. the left and right-handed fermions transform differently under the gauge
symmetry, and therefore they should be massless. However, a miracle happens here: the Higgs mechanism with a 
minimal Higgs doublet induces the mass for the SM fermions too. 
Fermion masses are obtained from the Yukawa Lagrangian that includes the coupling of the fermion doublet $F_L$  with the Higgs doublet
$(\Phi)$ and the right-handed fermion singlet $f_R$, which  for the leptons and quarks take the form:
\begin{equation}
{\cal{L}}_{Yukawa}=y_{e} \bar{L} \phi e_{R} + y_{d} \bar{Q}_L \phi d_R + y_u \bar{Q}_L (i\sigma_2\phi^*) u_R+h.c.
\end{equation}
After SSB, all the SM fermions acquire masses (except the neutrinos), which is given by: 
$m_f=y_f\frac{v}{2}$. The Yukawa Lagrangian includes the Higgs-fermion interactions too, which turn out to be
proportional to the fermion mass, i.e. $(hff)= \frac{m_f}{v}$.


\subsection{SM Higgs parameters: LEP and indirect constraints}

Within the SM, the Higgs sector includes only two parameters:
the (dimensional) quadratic mass term $\mu^2$ and the dimensionless quartic coupling
$\lambda$. Once SSB is implemented, these parameters can be traded by the Higgs vev ($v$) and the Higgs mass ($m_h$).
But what to expect for $\lambda$? Over the years, a cocktail of arguments was developed that helped to
constrain the Higgs mass, namely.

\begin{itemize}
\item {\bf{Unitarity:}} The processes involving gauge bosons and the Higgs, such as $WW$ scattering, should satisfy the
unitarity bounds, which also translate into a Higgs mass bound of $O(1)$ TeV \cite{Lee:1977eg,Chanowitz:1978uj}, 
unless the Higgs interactions become strong.
\item {\bf{Perturbativity:}} Given that the EW gauge sector is weakly-interacting, one would expect that
the Higgs interactions should also be weak, which would imply: $\lambda \simeq O(1)$, and then the Higgs mass
was to be expected to be of order of the electroweak scale, i.e. $m_h = O(v)$ \cite{Cabibbo:1979ay,Lindner:1985uk}.
\item {\bf{Vaccum stability:}} A lower bound on the Higgs mass was also derived from requiring that the radiative corrections do not 
make the potential to develop instabilities, but this one was quickly overcome after LEP bounds on the Higgs mass.. 
\item {\bf{Radiative corrections:}} It is also possible to derive constraints on the Higgs mass by considering the indirect effects of 
the Higgs on the  EW observables.  This could be done by using only the corrections to the $\rho$ parameter, as it was pioneered 
by Veltman \cite{Veltman:1977fy}.  With the precision reached at LEP (and Tevatron) it was possible to refine this analysis and use 
the complete set of precision tests, up to the point that it was known that the Higgs mass should be at the reach of LHC. 
\end{itemize}

A graphic summary of these constraints was presented in ref. \cite{Kolda:2000wi}, which we show in figure 2 (left). 
Thus,  already by 2000, it was known that the favored Higgs mass range was $110 < m_h< 180$ GeV.
Fig. 2 (right), from ref. \cite{Erler:2000cr}, shows that the probability distribution for the Higgs mass, which is constrained 
by radiative corrections
to lay in the range $110 < m_h< 130$ GeV.

\begin{figure}[t]
\begin{center}
\includegraphics[width=7.7cm, height=6.3cm]{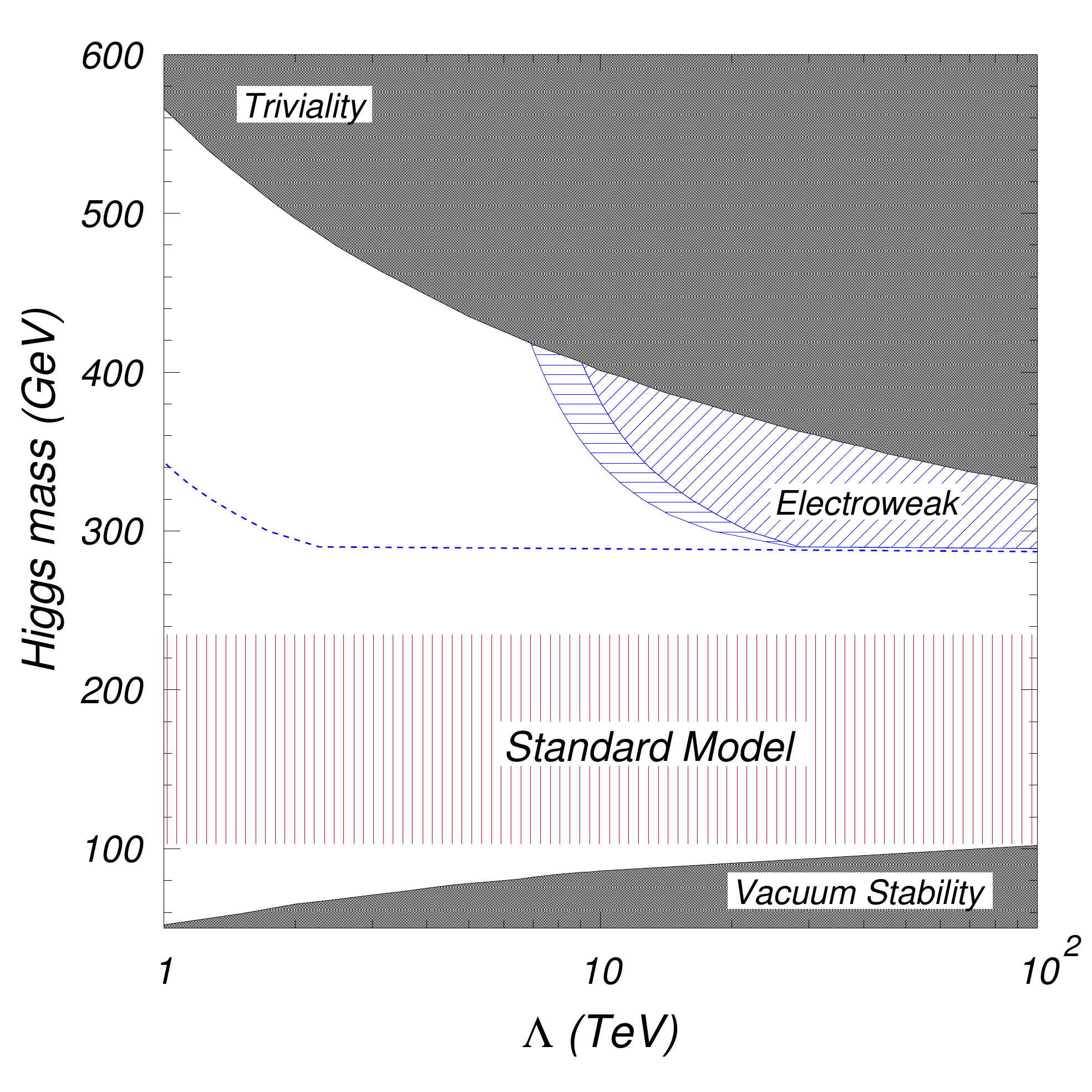}
\includegraphics[width=7.7cm, height=6.3cm]{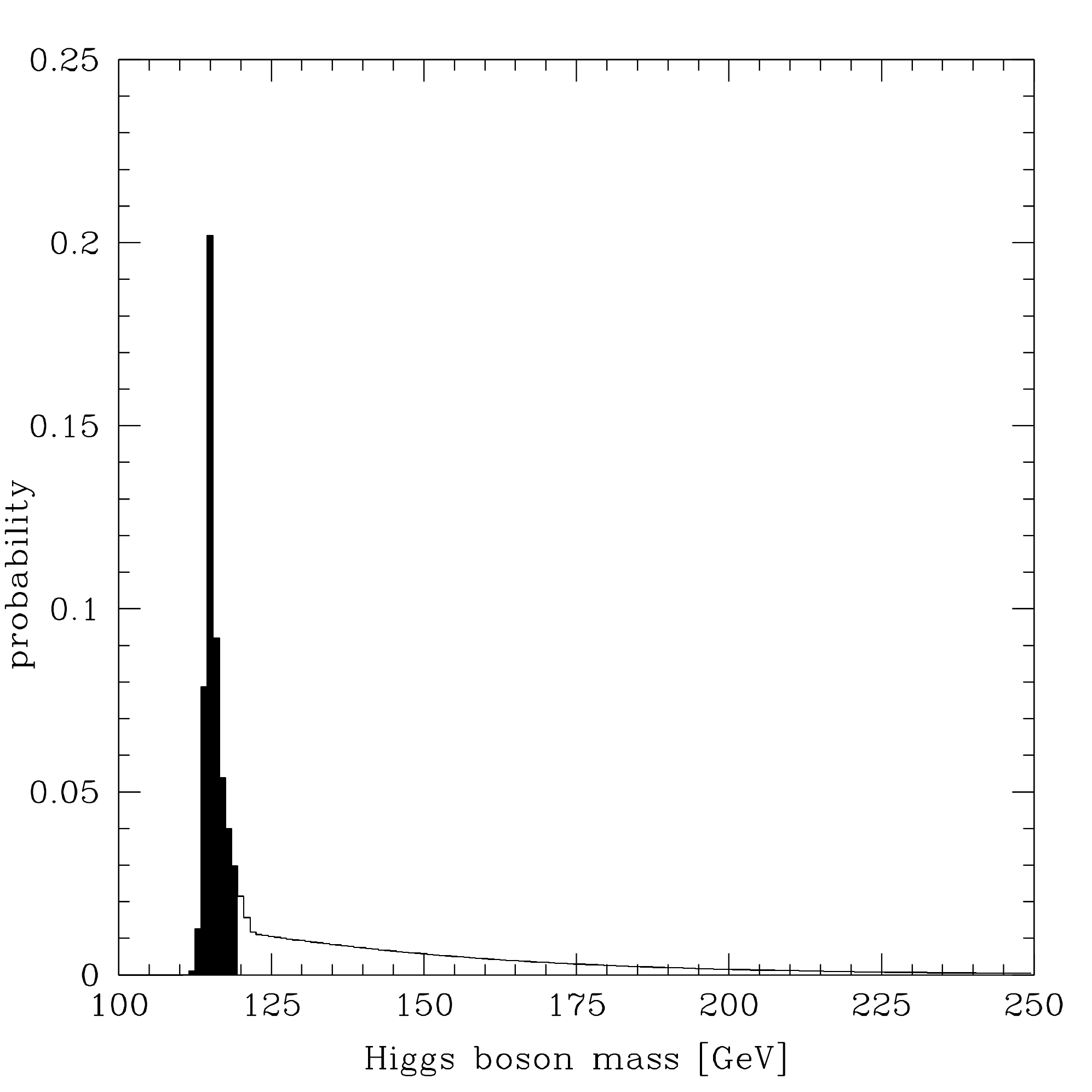}
\caption{Higgs mass limts from theoretical considerations (left) and EWPT (right) . 
Figure 2- -left is from \cite{Kolda:2000wi}, and figure 2-right is  from \cite{Erler:2000cr} }
 \label{Fig:HiggsMassEWPT}
 \end{center}
\end{figure}

\section {Higgs phenomenology and the LHC }

As good as it can be, the theoretical bounds on the Higgs mass, including the analysis of radiative corrections, only provided 
indirect tests of the Higgs mass. To prove the existence of the Higgs particle, one must produce it through some reaction 
at some accelerator and to  detect its decay products in the particle detectors; this task started at LEP, continued at the Tevatron 
and finally it was completed at the LHC. The starting point for such analysis is to have the Feynman rules of the Higgs sector, 
which comes next.

\subsection{General Couplings and Feynman rules}

As mentioned in the introduction, in order to study the Higgs properties at the LHC it is both instructive and useful
to present a  parametrization of the Higgs couplings that is general enough, while at the same time it makes easy to 
reduce to the expressions to the SM limit, or some other popular SM extension, such as the 2HDM.

Thus, we shall start by describing the coupling of the scalar boson ($h$) with Vector bosons $W^\pm, \, Z$. In this case we can write
the interaction Lagrangian with terms of dimension-4, consistent with Lorentz symmetry and derivable from a renormalizable
model, as follows:

\begin{equation}
 {\cal{L}}_V = \kappa_W gm_W h W^{+\mu} W^-_{\mu} + \kappa_Z \frac{gm_Z}{2c_W} h Z^{\mu} Z_{\mu}
\end{equation}

When the Higgs particle corresponds to the minimal SM, one has $\kappa_W=\kappa_Z= 1$, while values 
$\kappa_W=\kappa_Z \neq 1$ arise in models with 
several Higgs doubles, which respect the so-called custodial symmetry.

On the other hand, the interaction of the Higgs with fermion $f$, with $f=q$ or $l$ for quarks and leptons,
respectively, can also be written in terms of a dimension-4 Lagrangian that respects Lorentz invariance,
namely:

\begin{equation}
 {\cal{L}}_f =h \bar{f}_i ( S_{ij} + i \gamma_5 P_{ij}) f_j 
\end{equation}

The CP-conserving and CP-violating factors $S_{ij}$, $P_{ij}$, which include the flavor physics we are interested in,
 are written as:

\begin{eqnarray}
 S_{ij} &=& \frac{g m_i  }{2m_W} c_{f} \delta_{ij} + \frac{g  }{2} d_{f} \eta_{ij} \\
 P_{ij} &=& \frac{g m_i  }{2m_W} e_{f} \delta_{ij} + \frac{g  }{2} g_{f} \eta'_{ij}
\end{eqnarray}

Within the SM $c_f=1$ and $d_f = e_f = g_f =0$, which signals the fact that
within the SM the Higgs-fermion couplings are CP-conserving and  flavor diagonal, i.e. no FCNC scalar interactions. 
However, as we shall see next, the LHC analysis has been reported
by considering $c_f \neq 0$ (but with $d_f = e_f = g_f =0$).
These factors take into account the possibility that the Higgs boson is part of an enlarged
Higgs sector, and the fermion masses coming from more than one Higgs doublet. As we shall discuss in the next sections,
for specific multi-Higgs models, the explicit dependence
on the Yukawa matrix structure, i.e. flavor physics,  is contained in the factor $\eta_{ij}$, $\eta'_{ij}$.
On the other hand, when we have $e_f \neq 0$, but $d_f= g_f =0$, it indicates
that the Higgs-fermion couplings are CP violating, but still flavor diagonal.
In models with two or more Higgs doublets, it is possible to have both flavor changing scalar interactions 
($d_f, g_f  \neq 0 $)  and CP violation.

\subsection{Higgs decays} 

For the study of Higgs phenomenology, we need to know first the Higgs branching ratios, including its decay into the most relevant 
modes ($BR(h\to XX)$). For the mass range left by the analysis of EWPT, i.e. $m_{h} \approx 105 \rightarrow 130$, we know
that its main decays modes must be:  $h \rightarrow b \bar{b}, c \bar{c}, \tau^+ \tau^-, \gamma \gamma , ZZ, WW^{+}, gg$,
whose decay widths are presented in the literature (See for instance ref. \cite{Gunion:1989we}). The Branching ratio for
the mode $h\to XX$ is defined as:
\begin{equation}
 BR(h\to XX) = \frac{\Gamma (h\to XX) }{\Gamma_{total}}
\end{equation}
The total width $\Gamma_{total}$ is given by the sum over all the partial widths, and its value is about 
$10^{-3}$ GeV. 
These decays have been evaluated in the literature, including QCD and EW corrections.
Results for the corresponding branching ratios are shown in Fig. 3, from Ref. \cite{Dittmaier:2011ti}.
 We can see that the dominant mode is $h \rightarrow b \bar{b}$,
while $ h \to ZZ, WW^{+}$ come next, and $h \to \tau^+ \tau^-$ is also relevant. Although the loop decay into a photon pair
reaches a branching ratio of only about $2\times 10^{-3}$ for $m_h=120-130$ GeV, it has a great relevance because it provides a very clean signature. On the other hand, the loop decay into  gluon pair is important, not because it could be detected as a Higgs decay, but
because it serves for the Higgs production mechanism through gluon fusion.

\begin{figure}[t]\begin{center}
\includegraphics[width=9.7cm, height=7.3cm]{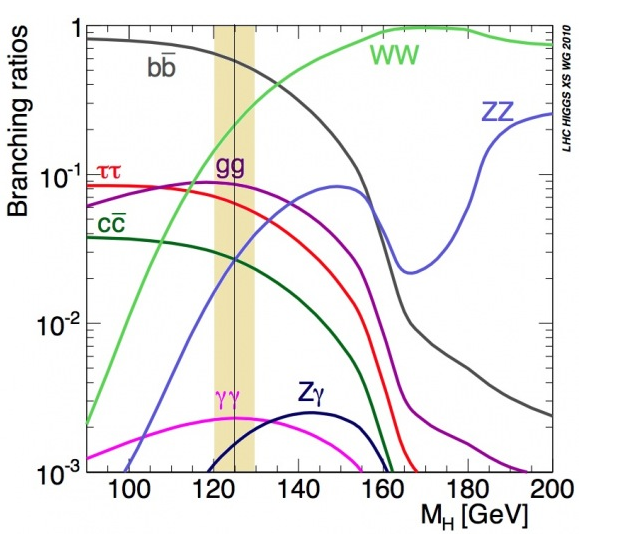}
\caption{Higgs couplings derived at LHC. Figure from Ref. \cite{Dittmaier:2011ti}.}
\label{Fig:SMHixBR}
\end{center}
\end{figure}

\subsection{Higgs search at LEP}

The clean environment of electron-positron colliders was used first to search for the Higgs boson. At the first stage of LEP,
which worked on the Z-pole, the Higgs was searched through the Z decay: $Z \to ff + h$; from the non-observation of such decay it
was possible to put the bound: $m_h \geq 80$ Gev.  Then, the second phase of LEP worked with a cm energy of 200 GeV, and 
then the search for the Higgs used the reaction (Bjorken mechanism):
$e^+ e^- \to Z+h$. Again, the absence of a signal resulted in the bound: $m_h \geq 200 - m_Z \simeq 110$ GeV.

In the late stages of LEP, with a cm energy of about 205 GeV, this bound was slightly extended, up to about 114 GeV. 
Just when the CERN plans marked that LEP must be closed, in order to start the LHC project, some debate arose
because there was a claim for the presence of a Higgs signal with $m_h =115$ Gev, which produced lots of enthusiasm among 
some experimentalist (who pledged for more time and energy for LEP) and also among some theorists, who quickly cooked models 
that could explain such mass value. The CERN director decided to close LEP and keep the plans for LHC, which a posteriori 
seems it was the best decision that  could be made.

\subsection{Higgs production at Hadron colliders}

 Although at a proton collider it is also possible to produce Higgs bosons with mechanism similar to the one used at LEP, 
 the Bjorken mechanism, 
 namely  with quarks and antiquarks colliding to produce the Higgs boson in association with a massive gauge boson 
 ($W$ or $Z$), it turns out that the dominant production mechanism is gluon fusion,
 despite the fact that it occurs at loop level (with top quarks circulating in the loop). This is so, because gluons carry the largest fraction of momentum from the colliding protons, and also because its parton distribution function provides with the
 largest luminosities. The production of the Higgs boson with a pair of heavy quarks (top mainly) also reaches a significant
 cross section. The resulting cross sections for these mechanisms in proton proton colliders, times the branching ratios of some 
 relevant modes, as a function of the cm energy,
 is shown in fig. 4  \cite{Dittmaier:2011ti}.

\begin{figure}[t]
\begin{center}
\includegraphics[width=7.7cm, height=6.3cm]{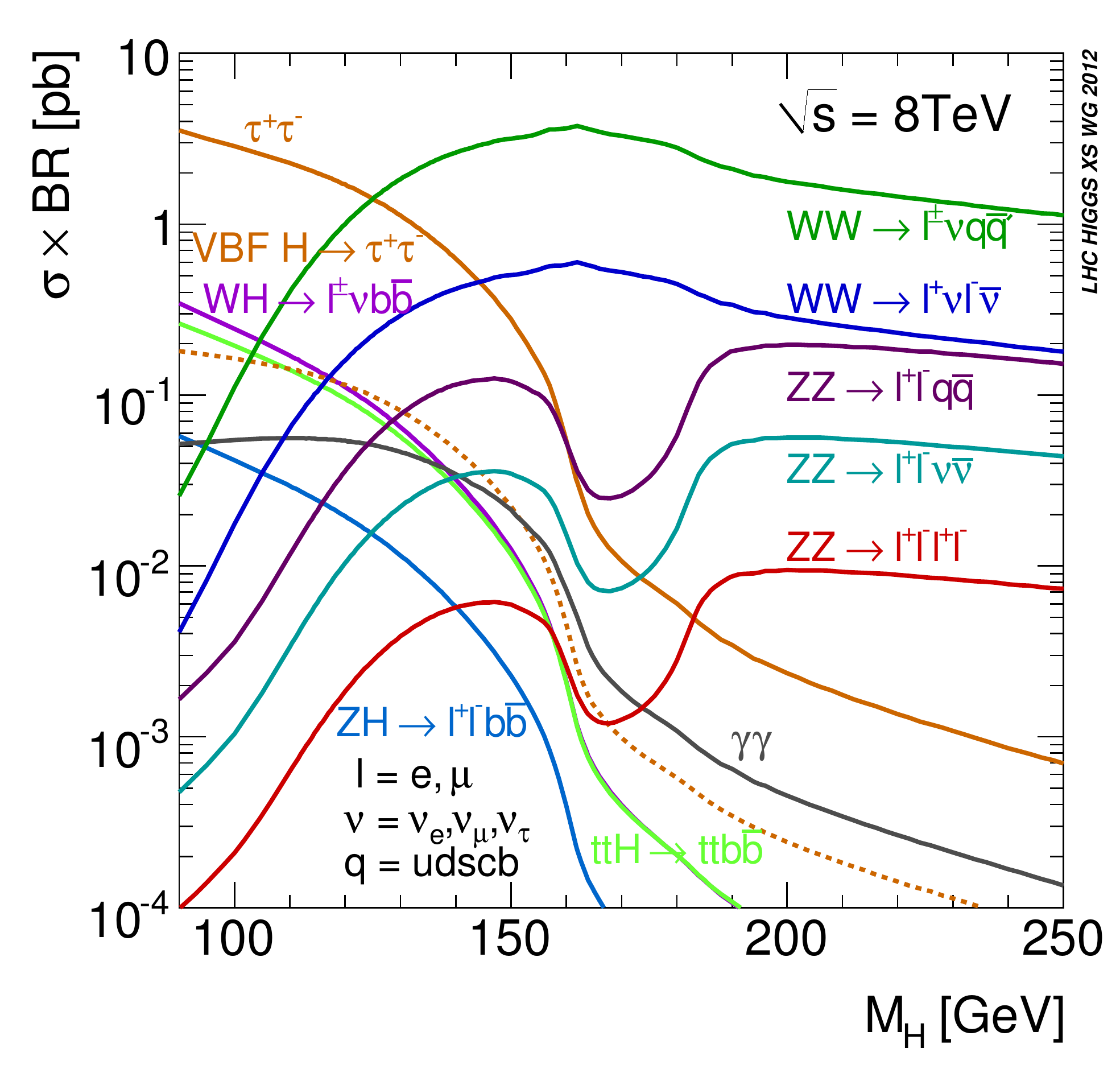}
\caption{Higgs cross section times BR for different modes, at LHC with Ecm=8 TeV. Figure from \cite{Dittmaier:2011ti} }
\label{Fig:HiggsXSBR-SM}
\end{center}
\end{figure}

\subsection{Current Higgs profile from LHC}

LHC started collecting data from pp collisions with cm energy of $\sqrt{S}=7$ TeV.
After some initial claims for a Higgs signal around 140 GeV, which was again quickly explained by some theory models, the 
Higgs became real when both collaborations (ATLAS and CMS)
announced on July 4th, 2012 that some events on the $\gamma\gamma$ and $ZZ *$ channels were observed, which
would correspond to a SM-like Higgs particle with mass $m_h= 125-126$ GeV. The observation of a resonance decaying into
$\gamma\gamma$ indicated that its spin must be either 0 or 2, in accordance with Yang theorem. An scalar, with spin-0, seemed 
the most natural explanation, which was reinforced by the second decay mode, namely the four-lepton signal, which could
be interpreted as coming from the decay $h \to ZZ^* \to llll$. Habemus Higgs!

As LHC continued to accumulate luminosity, more Higgs decay modes were observed, which confirmed that the particle 
observed by ATLAS and CMS was indeed a Higgs-like particle. 
The interpretation of the Higgs signal was further reinforced after LHC started taking data 
with higher energy ($\sqrt{S}=13$ TeV). Now the signal can be appreciated even by the public eye, for instance the data
from CMS on the four-lepton signal  shown in Fig. 5. (from  ref. \cite{Sirunyan:2017exp} ) shows clearly a bump in the invariant 
mass at 125 GeV (pink) ,  which stands clearly above the SM backgrounds.
There are now a variety of signals that have been measured, which seem all consistent with the SM, as it is shown in figure
6 (left) , from ATLAS (ref. \cite{Aad:2015gba}). The mass has been measured with better precision,
as it can be seen in figure 6 (right).

\begin{figure}[t]
\begin{center}
\includegraphics[width=9.7cm, height=7.3cm]{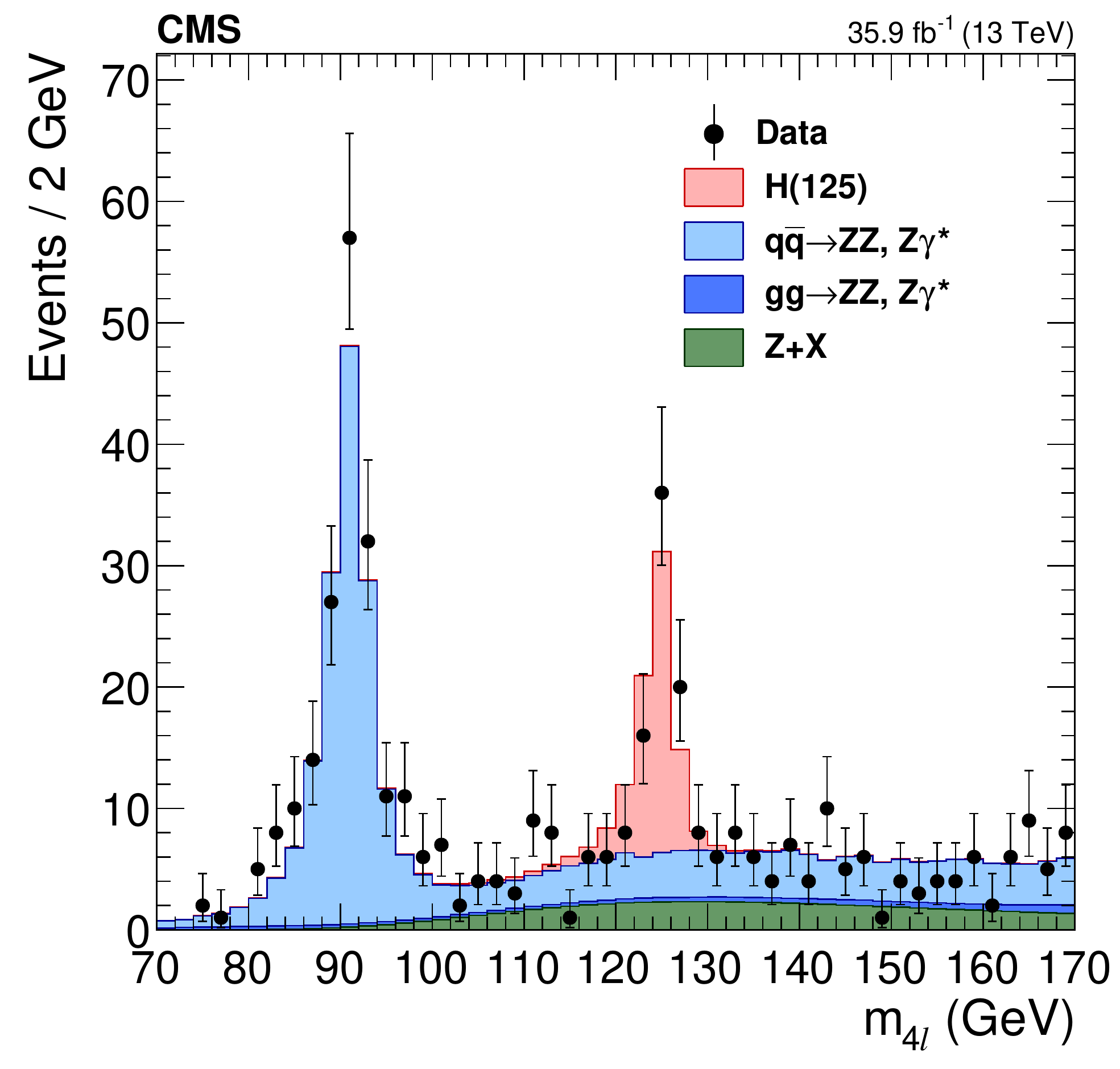}
\caption{Higgs signal from the decay $h\to ZZ^*$ from CMS. Figure from ref. \cite{Sirunyan:2017exp}}
\label{Fig:SMHix1}
\end{center}
\end{figure}

\begin{figure}[t]
\begin{center}
\includegraphics[width=7.7cm, height=6.3cm]{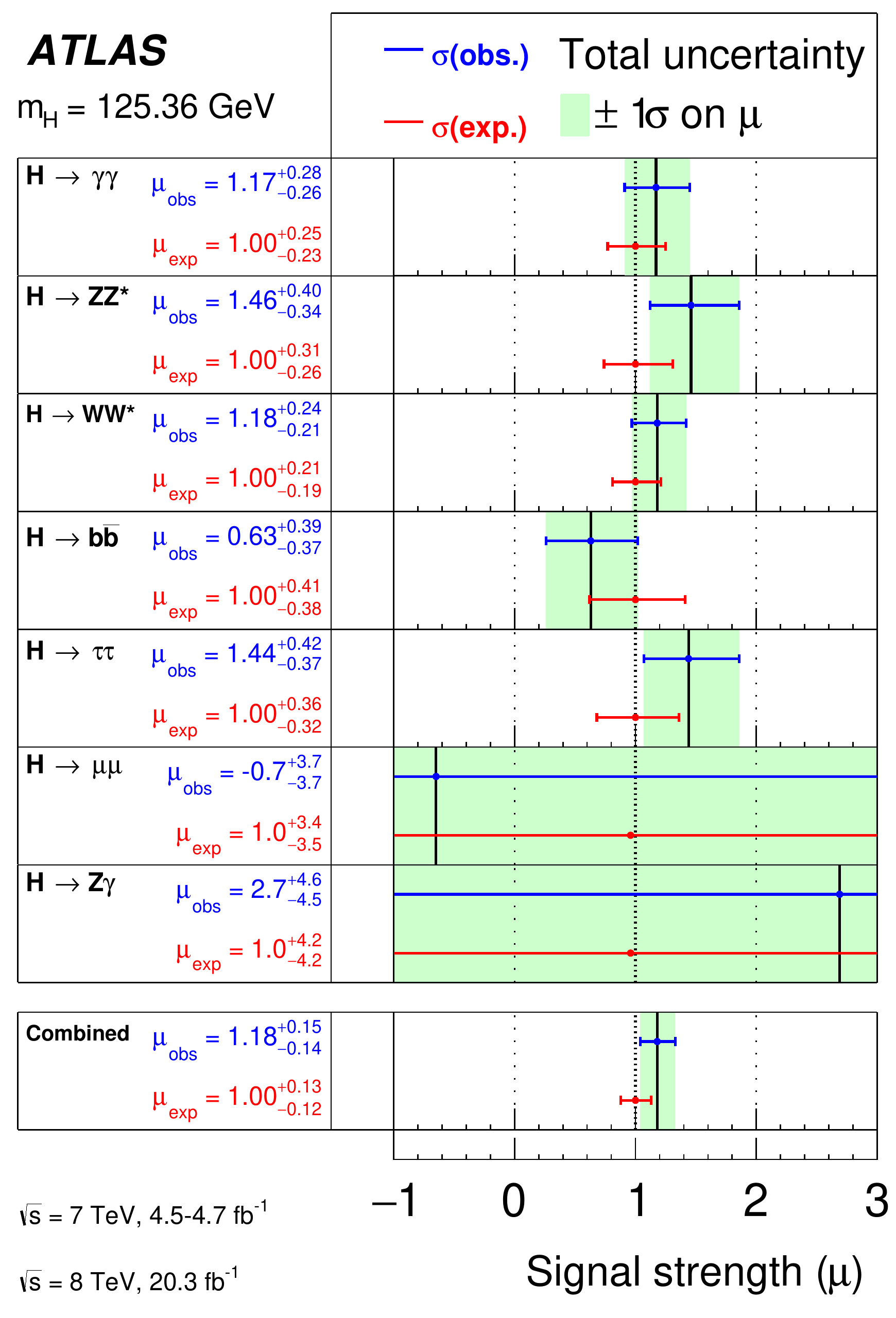}
\includegraphics[width=7.7cm, height=6.3cm]{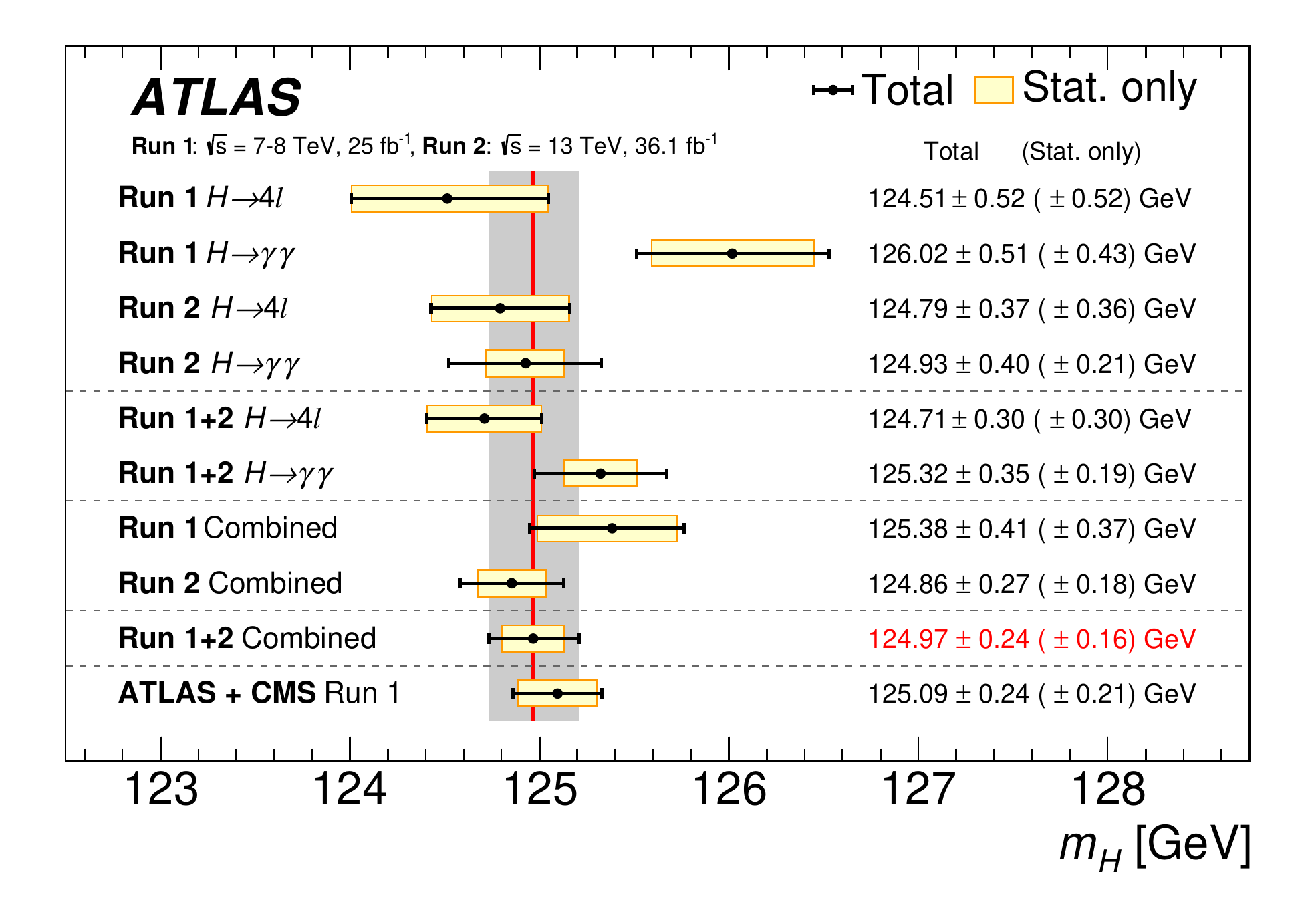}
\caption{Higgs signals (left) and mass summary (right) from ATLAS. Figure from ref. \cite{Aad:2015gba}}
\label{Fig:SMHix2}
\end{center}
\end{figure}

The current LHC data on Higgs production has been used to derive bounds on deviation from the SM
predictions for the Higgs couplings. For instance, figure 7 from Atlas Collaboration (from ref. \cite{Aaboud:2018wps}) 
shows the allowed deviations for the Higgs couplings with gauge bosons and fermions, 
 as parametrized by the constants $\kappa_V$ and $\kappa_F$. Thus, 
one can see that best fit lays quite close to the point $\kappa_V=\kappa_F=1$, which corresponds to the SM limit. 
However, small deviations are allowed within the 
precision reached at LHC.

\begin{figure}[t]
\begin{center}
\includegraphics[width=9.7cm, height=7.3cm]{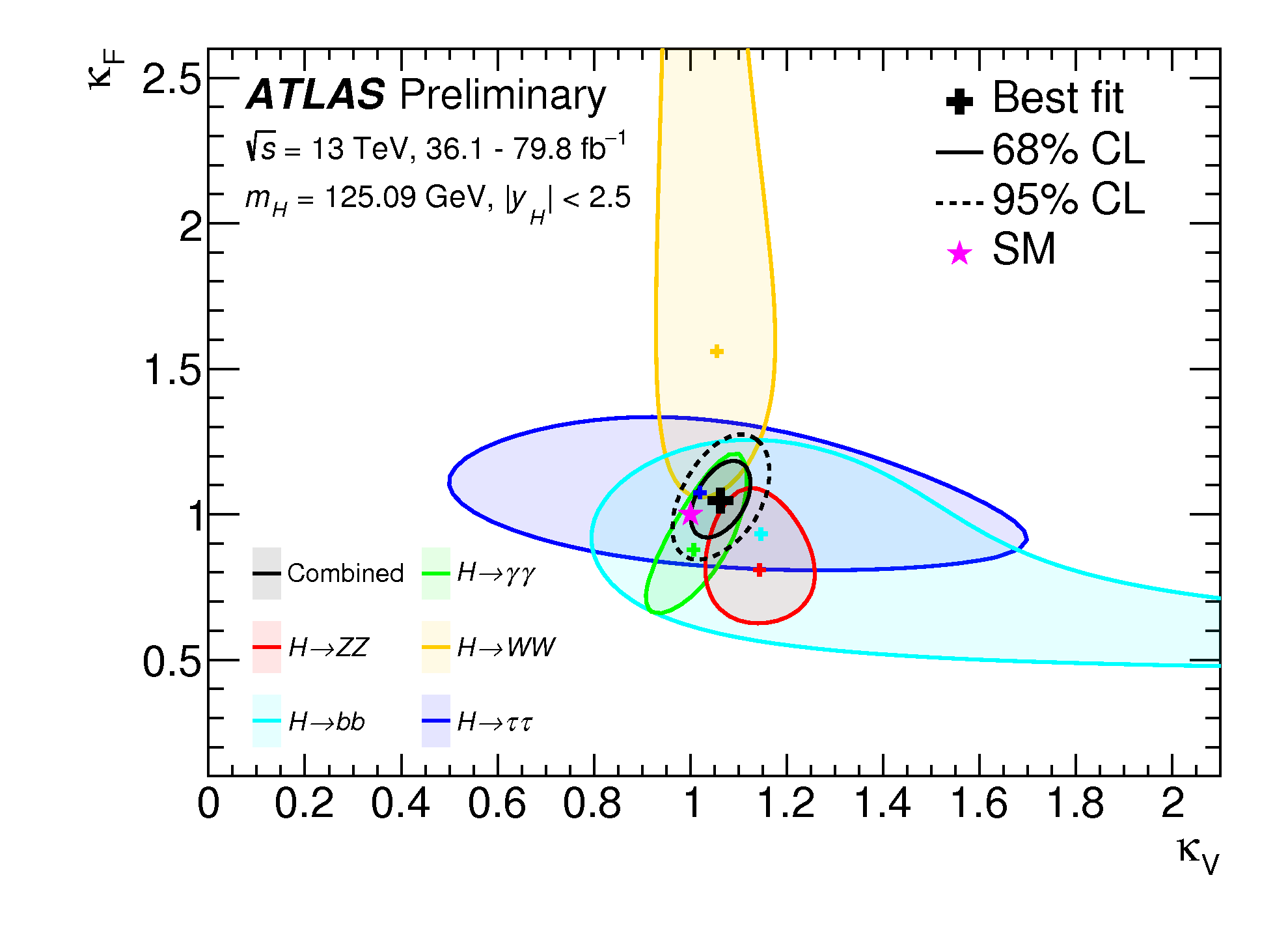}
\caption{Higgs couplings with vectors and fermions. Figure from ref. \cite{Aaboud:2018wps}}
\label{Fig:SMHix3}
\end{center}
\end{figure}


In essence, the Higgs particle couples to a pair of massive gauge bosons or fermions
with a strength proportional  to their masses.  So far, the LHC has tested only a few of them,
namely the Higgs couplings with the gauge bosons and the  heaviest  fermions of the third family. 
This is shown in figure 8, from both ATLAS (\cite{Aaboud:2018wps}) and CMS (\cite{Sirunyan:2018koj}) collaborations, 
where one can appreciate that the Higgs 
couplings lay on a straight line, modulo some small deviations for the Higgs coupling with $b\bar{b}$ 
(which is still consistent at one sigma); the mode $\mu\mu$  has not been detected yet. 
These results, allow some regions of parameter space for the so-called  
"private Higgs" hypothesis, where each fermion type gets its mass from a different Higgs doublet,
as it will be discussed briefly in our conclusions. Moreover, 
these constraints are obtained assuming SM-like pattern for the Higgs couplings, however when
one considers new physics, i.e. models beyond the SM
\cite{Branco:2011iw}, it is possible to have  non-standard Higgs couplings, 
including the flavor  and  CP-violating ones. These will be discussed in the coming sections.

\begin{figure}[t]
\begin{center}
\includegraphics[width=7.7cm, height=6.3cm]{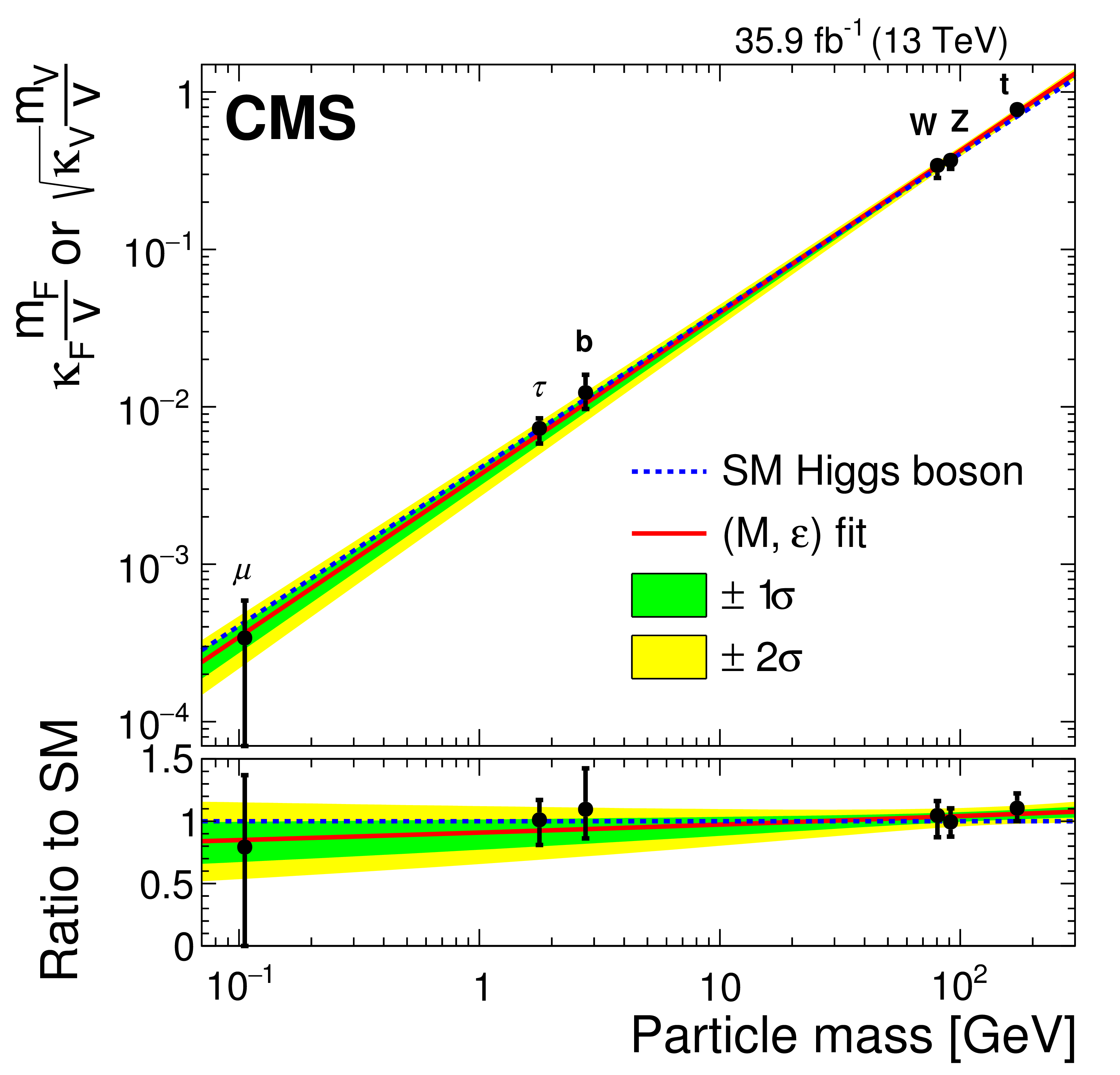}
\includegraphics[width=8.7cm, height=6.1cm]{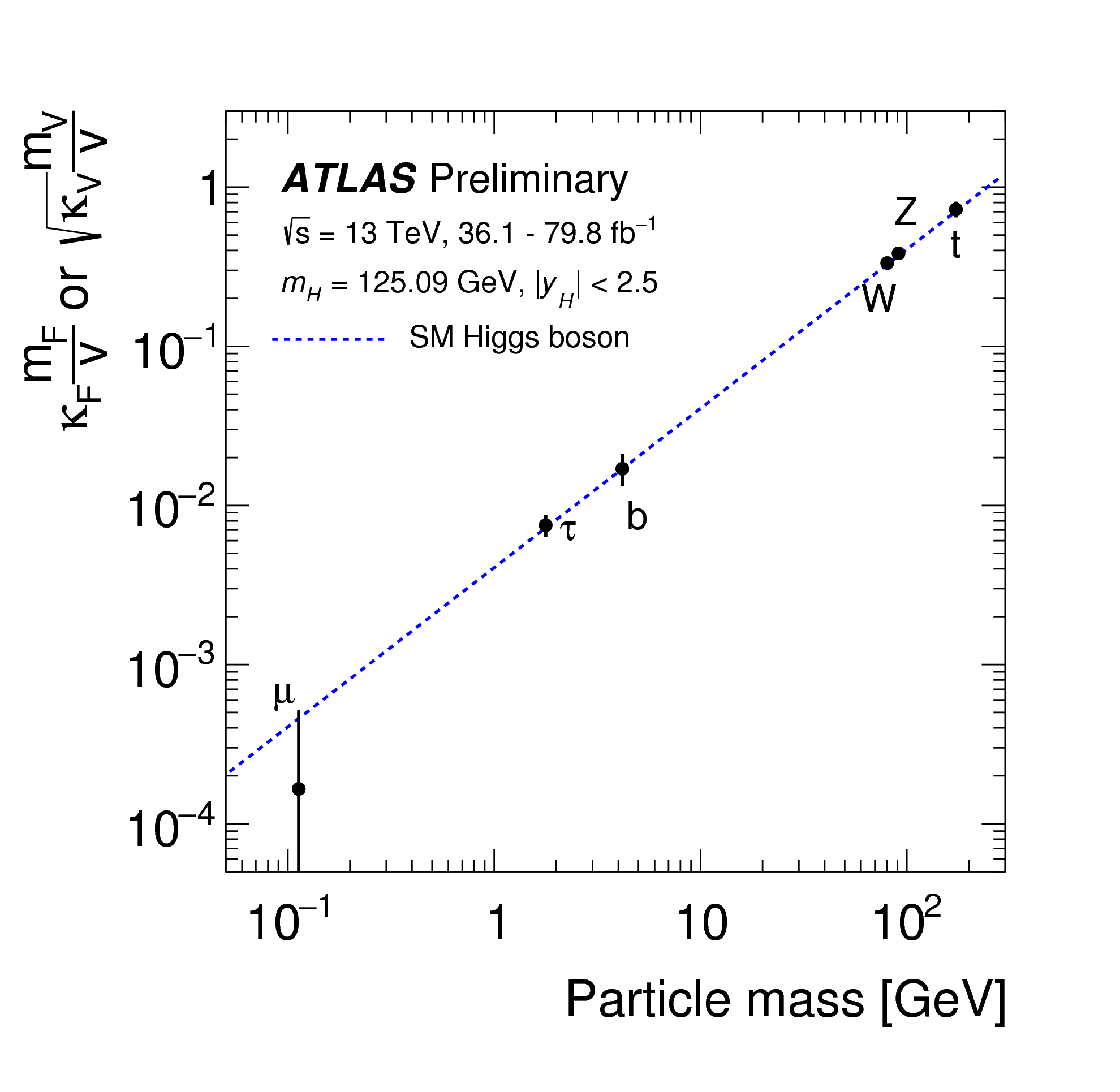}
 \caption{Higgs couplings as function of the mass from CMS (from \cite{Sirunyan:2018koj}) and ATLAS (from \cite{Aaboud:2018wps}).}
  \label{Fig:SMHix4}
 \end{center}
\end{figure}

\section{Beyond the SM Higgs sector:  SUSY, flavor  and Dark Matter}

The SM has several shortcomings that make us suspect that it is not a final theory.
For instance, the SM is unable to explain some of the most pressing
theoretical problems (unification, flavor, naturalness, etc) \cite{Arkani-Hamed:2016kpz}, as well as some 
cosmological data (DM, Dark energy, etc). In particular, finding a possible solution to the hierarchy problem suffered 
by  the SM Higgs boson, has  been the driving force  behind many of the proposals for extending the SM \cite{Pomarol:2012sb}. 

As we mentioned before, the SM Higgs particle couples to a pair of massive gauge boons or fermions, with an strength 
proportional to its mass. However,  so far LHC provided information on the essential Higgs properties, but this
information is based  only on a few of the Higgs couplings,
i.e. the ones with the heaviest SM fermions and $W,Z$. 
Then, some questions arise:

\begin{itemize}
\item Why is the Higgs mass light? i.e. of the order of the EW scale,
 \item Do the masses of all fermion types (up-, down-quarks and leptons) arise from a
       single Higgs doublet? or, are there more Higgs multiplets participating in the game?
 \item Are the Higgs couplings to fermions diagonal in flavor space?
 \item Is there any hope to measure the Higgs couplings with the lightest quarks and leptons?
\end{itemize}

As we shall see next, different answers to these questions arise when one extends the SM.
Many of those extensions often include a rich Higgs  spectrum, or predict deviations from the SM 
Higgs properties. These models could either be the realization of
elaborated theoretical constructions, or just examples of model building machinery; both of them are usefull 
at the minimum because they provides a systematic generation of new collider signals to search for, as we shall discuss next.
Thus, in order to test these extensions, it will be very important to study the Higgs couplings at LHC 
and future colliders, and compare with predictions from extended models. 

\subsection{Higgs and Supersymmetry}

One of the most widely studied extensions of the SM is Supersymmetry. 
This beautiful idea implies that for each fermions there is a bosonic superpartner, as they are related by a SUSY
transformation. The minimal implementation of such idea is the so called 
minimal SUSY extension of the SM (MSSM) (for a review see \cite{Haber:1984rc, Martin:1997ns}).
The MSSM contains 3 families of chiral superfields, which include the quarks and leptons as well
as their scalar superpartners, the squarks and sleptons. The gauge sector is described with the vector superfields
associated with $SU(3)_c \times SU(2)_L \times U(1)_Y$, and includes the usual gauge bosons, but also  their fermionic
superpartners, the gauginos. In order to give masses to both up- and down-type fermions, one must have two
Higgs doublets, described by chiral superfields, and thus one also has the Higgsinos, as the fermionic superpartners.
A discrete symmetry, $R-$ parity, must be imposed in order to respect lepton and baryon number conservation.
This $R$- parity implies that superpertners are produced by pairs, and the lightest superpartner must be stable,
which provides the seeds for a dark matter candidate, which was considered a great virtue of
SUSY models. As the experimental evidence shows, SUSY must be broken, and a realistic scheme is provided by
the Soft-breaking terms. Namely, SUSY is assumed to be broken in a Hidden sector, which is then transmitted to the
MSSM superpartners by a suitable mediator, which may involve gravity or gauge interactions, or anomaly mediation.

The search for the signals of the superparners has been one of the central goals of current 
and future colliders \cite{Kane:2006hd}.
However, the LHC has provided bounds on the new physics scale ($\Lambda$), either from the search for
new particles or from the effects of new interactions, with values
that are now entering into the multi-TeV range. This result is casting some doubts about
the theoretical motivations for new physics scenarios that were promoted assuming a mass scale 
of order TeV.  This is particularly disturbing for  
the concept of naturalness, and its supersymmetric implementation, since the bounds 
on the mass of superpartners are passing the TeV limit too \cite{Kane:2006hd}.
 However, SUSY is such a beautiful theory that it certainly desserves further work on its foundations
 and possible realization in nature.
 Thus, one has to wait for the next LHC stages, with higher energy and luminosity,
in order to obtain stronger limits, both on the search for new particles, such as heavier
Higgs bosons \cite{Arganda:2013ve,Arganda:2012qp,Chakraborty:2013si} or others, and for
precision tests of the SM properties.

\subsection{Higgs and flavor: A more flavored Higgs sector}

Although the Higgs boson has  diagonal couplings to the SM fermions at tree-level, 
it could mix with new particles that have a non-aligned flavor structure, and this permits to 
induce corrections to the diagonal Yukawa couplings and/or new flavor-violating (FV) Higgs interactions.
Non-standard Higgs couplings are  predicted in many models of physics beyond the SM.
This could happen, for instance, in extensions of the SM that
contain additional scalar fields, which have non-aligned 
couplings to the SM fermions. When these fields mix with the Higgs boson, it is possible to
 induce new FV Higgs interactions.   This transmission of the flavor structure to
the  Higgs  bosons, was discussed in our earlier  work \cite{DiazCruz:2002er},
where we called it  {\it{ more flavored Higgs boson}}.

This  occurs, for instance, in Froggatt-Nielsen type models, where one includes
a SM singlet (Flavon) that participates in the generation of the Yukawa hierarchies. 
This singlet mixes with the Higgs doublets, and induces FV Higgs couplings at tree-level.
The phenomenology for the mixing of the SM Higgs doublet with a flavon singlet has 
been studied in ref. \cite{Dorsner:2002wi,Arroyo-Urena:2018mvl}.
Mixing of SM fermions with exotic ones, could also 
induce FV Higgs couplings at tree-level  \cite{Langacker:1988ur}.
 On the other hand, within supersymmetric models, like the MSSM, the sfermion/gauginos can  have 
non-diagonal couplings to Higgs bosons and SM fermions. Then, FV couplings of the
Higgs with SM fermions could be induced at loop-level \cite{DiazCruz:2002er}.

In the next section, we shall discuss what is probably the simplest model that contains
 a more flavored Higgs sector, namely the general two-Higgs doublet model
  \cite{Branco:2011iw}.  

\subsection{Higgs and Dark Matter}

Understanding the nature of dark matter is another aspect of physics beyond the SM, which  motivates
extensions of the SM. The dark matter could interact with ordinary mater 
through the Higgs boson, an scenario called the Higgs portal. It could also happen that dark matter is
contained within an extension of the Higgs sector, which could include 
a  singlet , an extra (Inert) Higgs doublet, a mixture of them  or even a 
higher-dimensional multiplet \cite{Ma:2014rua}. DM candidates can also arise in composite Higgs
models \cite{DiazCruz:2007be}.

The case when DM candidate is contained in an extra Higgs doublet is known as the IDM \cite{Barbieri:2006dq}, 
which has been studied extensively in the literature \cite{Ma:2006km,LopezHonorez:2006gr,Krawczyk:2013jta}.
Including dark matter candidate with extra sources of CP violation, has motivated the constructions of models
with inert doublet and singlet \cite{Bonilla:2014xba}.

Within those models,  one can accommodate a neutral and stable scalar particle, which has the right mass and couplings 
to satisfy the constraints from DM relic density and direct/indirect searches. It also implies modifications of the SM-like
Higgs couplings, which must mass the LHC Higgs constraints. The heavy Higgs spectrum provides additional signatures
of these models, which could also be searched at future colliders.

\section{The  general 2HDM with Textures  and its limits }

One of the simplest proposals for physics Beyond the Standard Model, is the
so called Two-Higgs Doublet Model (2HDM),  which was initially studied in connection with
the search for the origin of CP violation \cite{Deshpande:1977rw}, and later on it was found to 
be connected with other theoretical ideas in particle physics, 
such as supersymmetry, extra dimensions \cite{Quiros:2006my} 
and strongly interacting systems \cite{Pomarol:2012sb, Aranda:2007tg}.
Models with extra Higgs doublets are safe regarding the $\rho$ parameter,  as they respect the
custodial symmetry which  protect the relation $\rho=1$.

Several possible realizations of the general 2HDM have been considered in the literature, 
which have been known as Type I, II and III.  The table 1 shows the Higgs assignments that
participate in the generation of the masses for the different types of fermions for the 2HDM's. 
Model I can have an exact discrete symmetry $Z_2$, which permits a possible dark
matter candidate coming from the $Z_2-$odd scalar doublet \cite{Ginzburg:2010wa}; within this
type I models, a single  Higgs doublet gives mass to the up, down quarks and leptons. 
The type II model \cite{Ginzburg:2004vp} assigns one doublet to each fermion type, $\Phi_1$ for leptons
and down-type quarks, and $\Phi_2$ for up-type quarks;
this type II model also arises in the minimal SUSY extension of the SM \cite{Gunion:1988yc}.
Both models of type I and II, posses the property of natural flavor conservation, as they respect 
 the Glashow-Weinberg Theorem \cite{Glashow:1976nt}, which  suffices to avoid Flavor Changing 
Neutral Currents (FCNC) mediated by the Higgs bosons, at the tree-level.

\begin{center}
\begin{table}
\begin{tabular}{|c|c|c|c|}
\hline Model type & Up quarks  & Down quarks & Charged leptons \\ \hline
2HDM-I & $\Phi_1$ &  $\Phi_1$ & $\Phi_1$  \\  \hline
2HDM-II & $\Phi_2$ & $\Phi_1$ & $\Phi_1$ \\  \hline
2HDM-X & $\Phi_2$ & $\Phi_2$ & $\Phi_1$ \\  \hline
2HDM-Y & $\Phi_2$ & $\Phi_1$ & $\Phi_2$ \\  \hline
2HDM-III & $\Phi_{1,2}$ & $\Phi_{1,2}$ & $\Phi_{1,2}$  \\  \hline
\end{tabular}
\caption{Higgs interaction with fermions for the different 2HDM types.}
\label{table-0}
\end{table}
\end{center}

There are also other models discussed in the literature called type-X (also called lepton specific) 
and type-Y (also called flipped) . Although they can be consider
as variations of model II, it is important to mention  both of them, for completeness and also
because they have interesting phenomenology. The Higgs assignments are also shown in table 1.

On the other hand, within the most general version of the 2HDM, where both Higgs 
doublets could couple to all types of fermions, diagonalization of the full mass  matrix, does not imply that 
each Yukawa matrix is diagonalized, therefore FCNC can appear at tree level \cite{Glashow:1976nt}. 
In order to reproduce the fermion masses and mixing angles, with acceptable levels of  FCNC 
\cite{Fritzsch:1995nx, Branco:1999nb, Hall:1993ca}, one can assume that the Yukawa matrices 
have a certain texture form, i.e. with zeros in different elements, and some possible choices will be discussed in
the coming sections.

The general model has been previously referred to as the 2HDM of type III, although 
this has also been used to denote a different type of model \cite{Altmannshofer:2012ar}.
There are also other relevant sub-cases of the general 2HDM, such as  
the so-called Minimal Flavor violating 2HDM \cite{Buras:2010mh, D'Ambrosio:2002ex,Aranda:2012bv}. 
The Aligned model is another example of viable 2HDM \cite{Pich:2009sp}. 
Thus, in order to single out the two-Higgs doublet model with textures among this diversity, 
we proposed to baptize it as the 2HDM-Tx \cite{Arroyo:2013tna}.

\subsection{Yukawa textures: parallel and complementary}
 
The initial construction of the 2HDM with textures \cite{Cheng:1987rs}, considered first
the specific form with six-zeros, and some variations with cyclic textures. In that case
a specific pattern of FCNC Higgs-fermion couplings of size $\frac{\sqrt{m_i m_j }}  {v}$ was found, which is  
known nowadays as  the Cheng-Sher ansatz. For this type of Higgs-fermion vertex it is possible to satisfy 
the FCNC bounds with  Higgs masses lighter than  O(TeV). The extension of the 2HDM-Tx with a four-zero 
texture was presented  in \cite{Zhou:2003kd, DiazCruz:2004tr}. 
The phenomenological consequences of these textures for Higgs physics (Hermitian 4-textures or non-hermitian  6-textures) were 
considered  in \cite{DiazCruz:2004pj}, while further phenomenological studies were presented in
\cite{Wu:2004kr, Li:2005rr, Carcamo:2006dp, DiazCruz:2010yq}.

 When the fermions ($f=U,D,L$) couples to both Higgs doublets, after SSB the mass matrix
is given by:
\begin{equation}
 M_f = \frac{v_1}{\sqrt{2}} Y^f_1 + \frac{v_2}{\sqrt{2}} Y^f_2
\end{equation}

Although the texture pattern is defined by considering the zeroes appearing in the fermion mass matrix $M_f$, we can have
several options for the textures of the Yukawa matrices, $Y_1$ an $Y_2$. In our work on the 2HDM-Tx  \cite{Arroyo:2013tna}, 
we considered several possibilities, namely:

\begin{itemize}
\item {\it{Parallel textures:}} In this case, we have that both $Y_1$ an $Y_2$ have the same texture pattern. For instance, if $M_f$ is 
of the four-texture type, then both $Y_1$ an $Y_2$ have the same four-texture.

\item {\it{Complementary textures:}} In this case, we have that both $Y_1$ an $Y_2$ have different texture pattern, without having
elements in common,  but in such a way that they produce $M_f$   of the given texture type. 

\item {\it{Semi-Parallel textures:}} In this case,  $Y_1$ an $Y_2$ have different texture patterns, but they have elements in
common, and in such a way that they produce $M_f$ with a given texture type.

\end{itemize}

For instance, a four-zero texture mass matrix is of the form:
\begin{eqnarray} 
{M_f} =  \left( \begin{array}{ccc}
0 & D & 0 \\
D^* & C & B \\
0 & B^* & A
\end{array} \right),\label{eq:mass}
\end{eqnarray} 

In the parallel case we have that both Yukawa matrices ($Y_i, \, i=1,2$) have the same form, namely:
\begin{equation}
Y_i = 
\begin{pmatrix}
0 & d_i & 0\\
d^*_i & c_i & b_i\\
0 & b^*_i & a_i \\
\end{pmatrix},
\end{equation}

An example of complementary case, where $Y^f_1$  and ${Y}^f_2$ (which only has a non-zero 33 entry)
combine to produce a mass matrix with four-zero texture,  is the following:
\begin{equation}
Y_1 = 
\begin{pmatrix}
0 & d_1 & 0\\
d^*_1 & c_1 & b_1\\
0 & b^*_1 & 0 \\
\end{pmatrix},
\hspace{5mm}
Y_2 = 
\begin{pmatrix}
0 & 0 & 0\\
0 & 0 & 0\\
0 & 0 & a_2 \\
\end{pmatrix}.
\end{equation}

Finally, an example of semi-parallel textures is the following:
\begin{equation}
Y_1 = 
\begin{pmatrix}
0 & d_1 & 0\\
d^*_1 & c_1 & b_1\\
0 & b^*_1 & a_1 \\
\end{pmatrix},
\hspace{5mm}
Y_2 = 
\begin{pmatrix}
0 & 0 & 0\\
0 & 0 & 0\\
0 & 0 & a_2 \\
\end{pmatrix},
\end{equation}
in this example, $Y^f_1$ has 
 a four-zero texture, while ${Y}^f_2$ has only a non-zero 33 entry.

The detailed study of these Yukawa matrices, including its diagonalization, and the phenomenological
consequences, was presented in our ref. \cite{Arroyo:2013tna}. Next, we shall derive
the Yukawa lagrangian of the general 2HDM-III, and we shall employ those textures when considering
 the 2HDM-Tx.

\subsection{The Yukawa lagrangian}

Within the general Two-Higgs doublet model, each Higgs doublet couples to a given fermion of type $f$ through the
Yukawa matrices $Y^f_1$ and $Y^f_2$, which combine after SSB to produce a fermion mass matrix with some texture.
We shall write the Yukawa Lagrangian in the 2HDM-III for the quark sector,  following 
the notation from \cite{DiazCruz:2004pj}, namely:

\begin{equation}
 L=  Y^u_1 
 { \bar{Q} }^0_{L} {\hgtt} u_{R}^0  + {{\yyuu}} \bar{Q}_{L}^0 {\hgtt} u_{R}^ 0 + 
 {{\yd}} \bar{Q}_{L}^0 {\hggs} d_{R}^0 + {Y^d_2} \bar{Q}_{L}^0 {\hggss} d_{R}^0 + h.c.
\end{equation}

where the quark doublets, quark singlets and Higgs doublets are written as:
	
\begin{eqnarray}
\nn Q_L^0 &=& \left(
  \begin{matrix}
u_L \\[2mm]
d_L \\
  \end{matrix}
\right),
\hspace{3mm}
\overline{Q}_L^0 = \left( \overline{u}_L,\overline{d}_L \right), \\
 {\hggs} &=&  \left(
  \begin{matrix}
\phi_1^+ \\[2mm]
\phi_1^0 \\
  \end{matrix}
\right), 
\hspace{3mm}
 {\hggss} = \left(
  \begin{matrix}
\phi_2^+ \\[2mm]
\phi_2^0 \\
  \end{matrix}
\right), 
\hspace{3mm} 
\end{eqnarray}
And similarly for the leptons. We have defined the conjugate doublets as:
$\tilde{\Phi}_j   =   i \sigma_2 {\Phi}^{*}_j = ( {\phi_j^0}^* , -\phi_j^- )^T $,
with $\phi_j^0 =( \varphi_i +i \chi_i)/{\sqrt{2}} $.

On the other hand, the CP-properties of the Higgs spectrum depend on the parameters of the Higgs
potential. For the 2HDM, this has been studied extensively \cite{Accomando:2006ga}, using either
explicit constructions \cite{Ginzburg:2004vp}, or employing the basis-independent conditions
that classify whether the vacuum respects CP,  it violates explicitly CP or  CP is spontaneously broken
\cite{Davidson:2005cw,Gunion:2005ja,Maniatis:2007vn,Ferreira:2010jy}. 
Here we shall follow the methods and notation from ref.  \cite{Haber:2010bw}.
Thus, the Higgs mass eigenstates ($H_i$) are obtained by the orthogonal rotations,
as follows:

\begin{equation}
\left(
\begin{array}{c}
\varphi _{1} \\
\varphi _{2} \\
\chi _{1} \\
\chi_{2}%
\end{array}
\right) =R\left(
\begin{array}{c}
H_{1} \\
H_{2} \\
H_{3} \\
H_{4}%
\end{array}%
\right) .  \label{higgs3}
\end{equation}
We identify $H_4=G^0$ as the neutral pseudo-Goldstone boson.

For the case with a CP-conserving Higgs potential, one has that $H _{4}= \cos \beta \chi_1  + \sin \beta \chi_2$ is the Goldstone boson,
where $\tan\beta=v_2/v_1$. In this case the couplings of the CP-even Higgs bosons, $h$ and $H$, with $WW$ or $ZZ$
bosons are given by $g_{hVV}= \sin(\beta-\alpha) g^{sm}_{hVV}$ and $g_{HVV}=\cos(\beta-\alpha) g^{sm}_{hVV}$, 
respectively.

The matrix $R$ can be used to express  the Higgs doublets in terms of the physical Higgs mass eigenstate,
 as follows:
\begin{equation}
\Phi _{1}=\left(
\begin{array}{c}
G^{+} \cos \beta - H^{+} e^{i\xi} \sin\beta \\
\frac{v  }{\sqrt{2}}\cos \beta + \frac{1}{\sqrt{2}}\sum_{r=1}^4 \left(
q_{r1} \cos \beta +q_{r2}e^{-i(\theta _{23}+\xi)} \sin\beta \right) H_{r}%
\end{array}%
\right) ,  \label{phys-eigen-A}
\end{equation}

and

\begin{equation}
\Phi _{2}=\left(
\begin{array}{c}
G^{+} e^{i\xi} \sin\beta + H^{+} \cos\beta \\
\frac{v  }{\sqrt{2}} e^{i\xi} \sin\beta +\frac{1}{\sqrt{2}}\sum_{r=1}^4 \left(
q_{r1} e^{i\xi} \sin\beta  +q_{r2}e^{-i\theta _{23}} \cos\beta \right) H_{r}%
\end{array}%
\right) ,  \label{phys-eigen-B}
\end{equation}


The values of $q_{ra}$ are shown in table
\ref{table-2}, they are written as combination of the $\theta_{ij}$, 
which are the mixing angles appearing in the rotation matrix
that diagonalize the mass matrix for neutral Higgs.

\begin{center}
\begin{table}
\begin{tabular}{|c|c|c|}
\hline $r$ & $q_{r1}$ & $q_{r2}$ \\ \hline
1 & $\cos \theta_{12}\cos \theta _{13}$ & $-\sin \theta _{12}-i\cos \theta {12}\sin\theta_{13}$\\
2 & $\sin \theta _{12}\cos \theta _{13}$ & $\cos \theta _{12}-i\sin \theta {12}\sin\theta_{13}$ \\
3 & $\sin \theta _{13}$ & $i\cos \theta _{13}$ \\
4 & $i$ & 0 \\ \hline
\end{tabular}
\caption{Mixing angles for Higgs bosons which consider spontaneous
and explicit CPV.}
\label{table-2}
\end{table}
\end{center}

After substituting these expressions in the Yukawa Lagrangian one obtains the Higgs-fermion interactions
for the general 2HDM-III. They are written in terms of the quark mass matrices ($q=U,D$), which
 receive contributions from both vev's, namely:
$ M_q = \frac{v_1}{\sqrt{2}} Y^q_1 + \frac{v_2}{\sqrt{2}} Y^q_2$.
As we mentioned in the previous sub-section, the diagonal form of  the quark mass matrices ($\bar{M}_q$) is obtained 
by applying the bi-unitary transformations, namely: $\bar{M}_q = V^q_L M_q V^{q\dagger}_R$. 
Explicit expressions for these matrices are shown in \cite{Arroyo:2013tna}.
Thus,  the couplings of the neutral Higgs boson ($H_r$, $a=1,2,3$)
 with fermions of type f can be expressed in the form of Eq. 23, as follows:
\begin{equation}
 {\cal{L}}_f =                H^a \bar{f}_i ( S^{fr}_{ij} + i \gamma_5 P^{fr}_{ij}) f_j 
\end{equation}

For the up-type quarks, the factors  $S_{ij}^{ur}$ and $P_{ij}^{ur} $ are written as:

\begin{eqnarray}
S_{ijr}^{ur} &=&\frac{1}{2v} \bar{M}_{ij}^{U}\left( q_{k1}^{\ast
}+q_{k1}-\tan \beta \left( q_{k2}^{\ast }e^{i\left( \theta _{23}+\xi
\right) }+q_{k2}e^{-i\left(
\theta _{23}+\xi \right) }\right) \right)   \nonumber \\
&&+\frac{1}{2\sqrt{2}\cos \beta }\left( q_{k2}^{\ast }e^{i\theta _{23}}%
\widetilde{Y}_{2ij}^{U}+q_{k2}e^{-i\theta _{23}}\widetilde{Y}%
_{2ij}^{U\dagger }\right)   \label{sugral}
\end{eqnarray}

and

\begin{eqnarray}
P_{ij}^{ur} &=&\frac{1}{2v}\bar{M}_{ij}^{U}\left( q_{k1}^{\ast
}-q_{k1}-\tan \beta \left( q_{k2}^{\ast }e^{i\left( \theta _{23}+\xi
\right) }-q_{k2}e^{-i\left(
\theta _{23}+\xi \right) }\right) \right)   \nonumber \\
&&+\frac{1}{2\sqrt{2}\cos \beta }\left( q_{k2}^{\ast }e^{i\theta _{23}}%
\widetilde{Y}_{2ij}^{U}-q_{k2}e^{-i\theta _{23}}\widetilde{Y}%
_{2ij}^{U\dagger }\right).   \label{pugral}
\end{eqnarray}

Similarly, for the down-type quarks we find:

\begin{eqnarray}
S_{ij}^{dr} &=&\frac{1}{2v}\bar{M}_{ij}^{D}\left[ q_{k1}+q_{k1}^{\ast
}-\tan \beta \left( q_{k2}^{\ast }e^{i\left( \theta _{23}+\xi
\right) }+q_{k2}e^{-i\left(
\theta _{23}+\xi \right) }\right) \right]   \nonumber \\
&&+\frac{1}{2\sqrt{2}\cos \beta }\left( q_{k2}e^{-i\theta
_{23}}Y_{2}^{D}+q_{k2}^{\ast }e^{i\theta
_{23}}\widetilde{Y}_{2}^{D\dagger }\right)   \label{sdgral}
\end{eqnarray}

and

\begin{eqnarray}
P_{ij}^{dr} &=&\frac{1}{2v}\bar{M}_{ij}^{D}\left[ q_{k1}-q_{k1}^{\ast
}+\tan \beta \left( q_{k2}^{\ast }e^{i\left( \theta _{23}+\xi
\right) }-q_{k2}e^{-i\left(
\theta _{23}+\xi \right) }\right) \right]   \nonumber \\
&&+\frac{1}{2\sqrt{2}\cos \beta }\left( q_{k2}e^{-i\theta _{23}}\widetilde{Y}%
_{2}^{D}-q_{k2}^{\ast }e^{i\theta _{23}}\widetilde{Y}_{2}^{D\dagger}
\right). \label{pdgral}
\end{eqnarray}

This Lagrangian is the most general one, which includes the possibility to have 
i) Deviations from the SM (diagonal) Higgs couplings  ii) New flavor violating Higgs couplings
and iii) New sources of CP violation, coming either from the Higgs potential or the Yukawa matrices.
Further simplifications can be obtained when one assumes that the Yukawa matrices are hermitic,
i.e. $ \widetilde{Y}_{2}^{Q} = \widetilde{Y}_{2}^{Q\dagger}$  (with $Q=U,D$) or when the Higgs potential
is CP conserving.
	
Next we shall present some limiting cases for the couplings of Higgs $H^r$ ($r=1,2,3$)  with fermions of type f , 
expressed in terms of the factors $S^{fa}_{ij}$, $P^{fa}_{ij}$, 
which in turn can be written as:
\begin{eqnarray}
 S^{fr}_{ij} &=& \frac{g m_i  }{2m_W} c_{fr} \delta_{ij} + \frac{g  }{2} d_{fr} \eta^{f}_{ij} \\
 P^{fr}_{ij} &=& \frac{g m_i  }{2m_W} e_{fr} \delta_{ij} + \frac{g  }{2} g_{fr} \eta^{'f}_{ij}
\end{eqnarray}

As it is discussed in detail in ref. \cite{DiazCruz:2004tr}, when one assumes parallel
 Yukawa matrices of the four-textur type, one can express those factors in terms of the quark masses, the
mixing angles and some flavor-dependent parameters. In fact, in that case the $\eta^{f}_{ij}$ 
parameters satisfy the Cheng-Sher ansazt and are universal (the same for all Higgs bosons), 
then it is possible, and convenient, to express them as follows:
\begin{equation}
\eta^f_{ij} =  \tilde{\chi}^f_{ij} \frac{ \sqrt{m_i m_j} }{m_W} 
\end{equation}
The parameters $\tilde{\chi}^f_{ij} $ can be constrained by considering all types of
low energy FCNC transitions, which produce the viable regions of parameter space. 

\subsection{The 2HDM-III with non-Hermitian Yukawa matrices and CPC Higgs potential}

In this case we shall consider that the Higgs sector is CP
conserving (CPC), while the Yukawa matrices are non-hermitian. Then,
without loss of generality, we can assume that $H_{3}=A$ is CP odd,
with $H_{1}=h $ and $H_{2}=H$ are CP even, then: $\cos \theta _{12}=\sin
\left( \beta -\alpha \right)$, $\sin \theta _{12}=\cos \left( \beta
-\alpha \right)$, $\sin \theta _{13}=0$, and $ e^{-i\theta
_{13}}=1$. The expressions for the neutral Higgs masses
eigenstates can be written now in terms of the angles $\alpha $ (which diagonalizes the CP-even
Higgs bosons) and $\beta $ ($\tan\beta=v_2/v_1$):

\begin{eqnarray}
\Phi _{a}^{0} &=&\frac{1}{\sqrt{2}}\left( v+h^{0}\sin \left( \beta
-\alpha
\right) +H^{0}\cos \left( \beta -\alpha \right) +iG^{0}\right) \widehat{v}%
_{a}  \nonumber \\
&&+\frac{1}{\sqrt{2}}\left( h^{0}\cos \left( \beta -\alpha \right)
-H^{0}\sin \left( \beta -\alpha \right) +iA^{0}\right)
\widehat{w}_{a}, \label{higgscpc}
\end{eqnarray}

where $a=1,2$, and for the CP-conserving limit $\widehat{v}_{a}$\ and $\widehat{w%
}_{a}$ have a vanishing phase $\xi =0$.
Here, the components of $\hat{v}_{a}$, $\hat{w}_{a}$ are given as:
\begin{equation}
\widehat{v}=\left(
\begin{array}{cc}
\hat{v}_{1}, & \hat{v}_{2}%
\end{array}%
\right) =\left(
\begin{array}{cc}
\cos \beta , & e^{i\xi }\sin \beta
\end{array}%
\right)   \label{v}
\end{equation}%
and%
\begin{equation}
\widehat{w}=\left(
\begin{array}{cc}
\hat{w}_{1}, & \hat{w}_{2}%
\end{array}%
\right) =\left(
\begin{array}{cc}
-e^{-i\xi }\sin \beta , & \cos \beta
\end{array}%
\right) .  \label{w}
\end{equation}%

Additionally, when one assumes a 4-texture for the Yukawa matrices,
the Higgs-fermion couplings further simplify as
$\widetilde{Y}_{2ij}^{U}=\chi _{ij}\frac{\sqrt{m_{i}m_{j}}}{v}$ .
Then, coefficient (\ref{sugral}) and (\ref{pugral}) for up sector
for $r=1,2,3$, and with $H_1=h$ being identified as the light SM-like Higgs boson, are shown in table III.

\begin{center}
\begin{table}
\begin{tabular}{|c|c|c|c|c|}
\hline 
Coefficient & $c_{ui}$ & $d_{ui}$ & $e_{ui}$  &  $g_{ui}$  \\ \hline
  $H_1$  & $\sin(\beta-\alpha)  - \tan\beta \cos(\beta-\alpha)$    & $\frac{\cos (\beta -\alpha) }{ \cos \beta }$ &  0 
  & $ \frac{\cos ( \beta -\alpha ) } { \cos \beta }$  \\  \hline
  $H_2$  & $\cos (\beta -\alpha )-\tan \beta \sin (\beta -\alpha )$ & $\frac{\sin(\beta -\alpha) }{ \cos \beta }$ &  0 
  & $ \frac{\sin(\beta -\alpha )}{ \cos \beta } $ \\  \hline
    $H_3$  & 0  & $\frac{1 }{ \cos \beta }$ &  $\tan\beta$
  & $ \frac{1}{ \cos \beta }$  \\  \hline
\end{tabular}
\caption{Coefficients for Higgs-Fermion couplings.}
\label{table-0}
\end{table}
\end{center}

The $\eta$ parameters are given as follows:
\begin{equation}
\eta^{u}_{ij} = \frac{\sqrt{m_{i}m_{j}}}{2\sqrt{2}v} \left( \chi _{ij}+\chi _{ij}^{\dag }\right)
\end{equation}

\begin{equation}
\eta^{u'}_{ij} =\frac{\sqrt{m_{i}m_{j}}}{2\sqrt{2}v} \left( \chi _{ij}-\chi^{\dag }_{ij}\right)
\end{equation}

Similar expressions are obtained for d-type quarks and leptons.

\subsection{The 2HDM-III with Hermitian Yukawa matrices and CPV Higgs }

In this case we assume the hermiticity condition for the Yukawa
matrices, but the Higgs sector could be CP violating.
For simplicity we shall consider that the Hermitic Yukawa matrices that obey a
four-texture form, but now CP is violated in the Higgs sector.

Then, one  obtains  the following expressions for the couplings of the neutral
Higgs bosons with the up-type quarks, namely:

\begin{equation}
S_{ijr}^{u}=\frac{1}{2v}M_{ij}^{U}\left[ q_{r1}^{\ast }+q_{r1}-\tan
\beta
\left( q_{r2}^{\ast }+q_{r2}\right) \right] +\frac{\sqrt{m_{i}m_{j}}}{2\sqrt{%
2}v\cos \beta }\chi _{ij}\left( q_{r2}^{\ast }+q_{r2}\right)
\label{sva}
\end{equation}

and
\begin{equation}
P_{ijr}^{u}=\frac{1}{2v}M_{ij}^{U}\left[ q_{r1}^{\ast }-q_{r1}-\tan
\beta
\left( q_{r2}^{\ast }-q_{r2}\right) \right] +\frac{\sqrt{m_{i}m_{j}}}{2\sqrt{%
2}v\cos \beta }\chi _{ij}\left( q_{r2}^{\ast }-q_{r2}\right)
\label{pva}
\end{equation}

similar expressions can be obtained for the down-type quarks and leptons, as well
as for the charged Higgs couplings.

\subsection{The 2HDM-III with Hermitic textures and CP-conservation (2HDM-Tx)}

In this case we assume the hermiticity condition for the Yukawa
matrices, and the Higgs sector is CP conserving.
For simplicity we shall consider that the Yukawa matrices obey a
four-texture form. 
Then,  the Higgs-fermion couplings take the following form.  For $r=1,2$ one gets,

\begin{eqnarray}
S_{ij1}^{u} &=&\frac{1}{v}M_{ij}^{U}\left( \sin (\beta -\alpha
)+\tan \beta
\cos (\beta -\alpha )\right)   \nonumber \\
&&-\frac{\chi _{ij}\sqrt{m_{i}m_{j}}}{\sqrt{2}v}\frac{\cos (\beta -\alpha )}{%
\cos \beta },
\end{eqnarray}%

\begin{equation}
S_{ij2}^{u}=-\frac{1}{v}M_{ij}^{U}\frac{\sin \alpha }{\cos \beta }+\frac{%
\chi _{ij}\sqrt{m_{i}m_{j}}}{\sqrt{2}v}\frac{\sin (\beta -\alpha
)}{\cos \beta },
\end{equation}%

For both Higgs bosons $H_1$ and $H_2$ one has: $P_{ij1}^{u}=P_{ij2}^{u}=0$
On the other hand, for $r=3$, one has $S_{ij3}^{u}=0$ and:
\begin{equation}
P_{ij3}^{u}=-i\frac{\chi _{ij}\sqrt{m_{i}m_{j}}}{\sqrt{2}v\cos \beta}.
\end{equation}

\subsection{The 2HDM of type I, II, X, Y:  CPC case}

In all these cases each fermion type (U,D,L) couple only with one Higgs doublet, as shown in table I. 
Now, we have that: $d_{fr}=g_{fr}=0$, while the non-zero coupling constants ($c_f$ for $h,H$ or 
$e_f$ for $A$),  are shown in table IV for  models I and II, while table V shows the corresponding
couplings for models X and Y. 

\begin{center}
\begin{table}
\begin{tabular}{|c|c|c|c|}
\hline Higgs boson      &  $c_u$  & $c_d$ &  $c_l$ \\ \hline
$h$ (I) &  $\cos\alpha / \sin\beta$ &   $\cos\alpha / \sin\beta$ & $\cos\alpha / \sin\beta$  \\  \hline
$h$ (II) &  $\cos\alpha / \sin\beta$ &   -$\sin\alpha / \cos\beta$ & -$\sin\alpha / \cos\beta$  \\  \hline
$H$ (I) &  $\sin\alpha / \sin\beta$ &   $\sin\alpha / \sin\beta$ & $\sin\alpha / \sin\beta$  \\  \hline
$H$ (II) &  $\sin\alpha / \sin\beta$ &   $ \cos\alpha / \cos\beta$ & $\cos\alpha / \cos\beta$  \\  \hline
$A$ (I) &   $\cot\beta$ & -$\cot\beta$ & -$\cot\beta$.  \\  \hline
$A$ (II) &   $\cot\beta$ & $\tan\beta$ & $\tan\beta$.  \\  \hline
\end{tabular}
\caption{Higgs interaction with fermions for flavor conserving 2HDM of types I and II.}
\label{table-0a}
\end{table}
\end{center}

\begin{center}
\begin{table}
\begin{tabular}{|c|c|c|c|}
\hline Higgs boson      &  $c_u$  & $c_d$ &  $c_l$ \\ \hline
$h$ (X) &  $\cos\alpha / \sin\beta$ &   $\cos\alpha / \sin\beta$ & -$\sin\alpha / \cos\beta$  \\  \hline
$h$ (Y) &  $\cos\alpha / \sin\beta$ &   -$\sin\alpha / \cos\beta$ & $\cos\alpha / \sin\beta$  \\  \hline
$H$ (X) &  $\sin\alpha / \sin\beta$ &   $\sin\alpha / \sin\beta$ & $\cos\alpha / \cos\beta$  \\  \hline
$H$ (Y) &  $\sin\alpha / \sin\beta$ &  $ \cos\alpha / \cos\beta$ & $\sin\alpha / \sin\beta$  \\  \hline
$A$ (X) &   $\cot\beta$ & -$\cot\beta$ & $\tan\beta$.  \\  \hline
$A$ (Y) &   $\cot\beta$ & $\tan\beta$ & -$\cot\beta$.  \\  \hline
\end{tabular}
\caption{Higgs interaction with fermions for flavor conserving 2HDM of types X and Y.}
\label{table-0a}
\end{table}
\end{center}

Then, the
Higgs couplings with fermions and gauge bosons in these models are determined by the mixing angles
$\alpha$ (which diagonalizes the neutral CP-even Higgs mass matrix) and $\tan\beta= v_2/v_1$.
For a quick test of Higgs couplings, one can rely on the Universal Higgs fit \cite{Giardino:2013bma}, where 
 bounds on the parameters $\epsilon_X$ are derived, they are defined as the (small) deviations of the 
Higgs couplings from the SM values, i.e. $g_{hXX}= g^{sm}_{hXX} (1 + \epsilon_X)$.
We find very convenient, in order to use these results and
get a quick estimate of the bounds, to write our parameters as: $\eta_X= 1 + \epsilon_X$. 
For fermions, the allowed values are:
$\epsilon_t= -0.21\pm 0.23$, $\epsilon_b= -0.19\pm 0.3$, $\epsilon_{\tau}= 0 \pm 0.18$.
However, specific tests of the mixing angles $\alpha$ and $\beta$, or combinations of them,  have been
presented by the  LHC collaborations; for instance the CMS constraints on the mixing angles $\alpha$ and $\beta$, for 2HDM-I and 2HDM-II, are shown in Figure 9 \cite{CMS:2016qbe}.

\begin{figure}[t]
\begin{center}
\includegraphics[width=7.7cm, height=6.3cm]{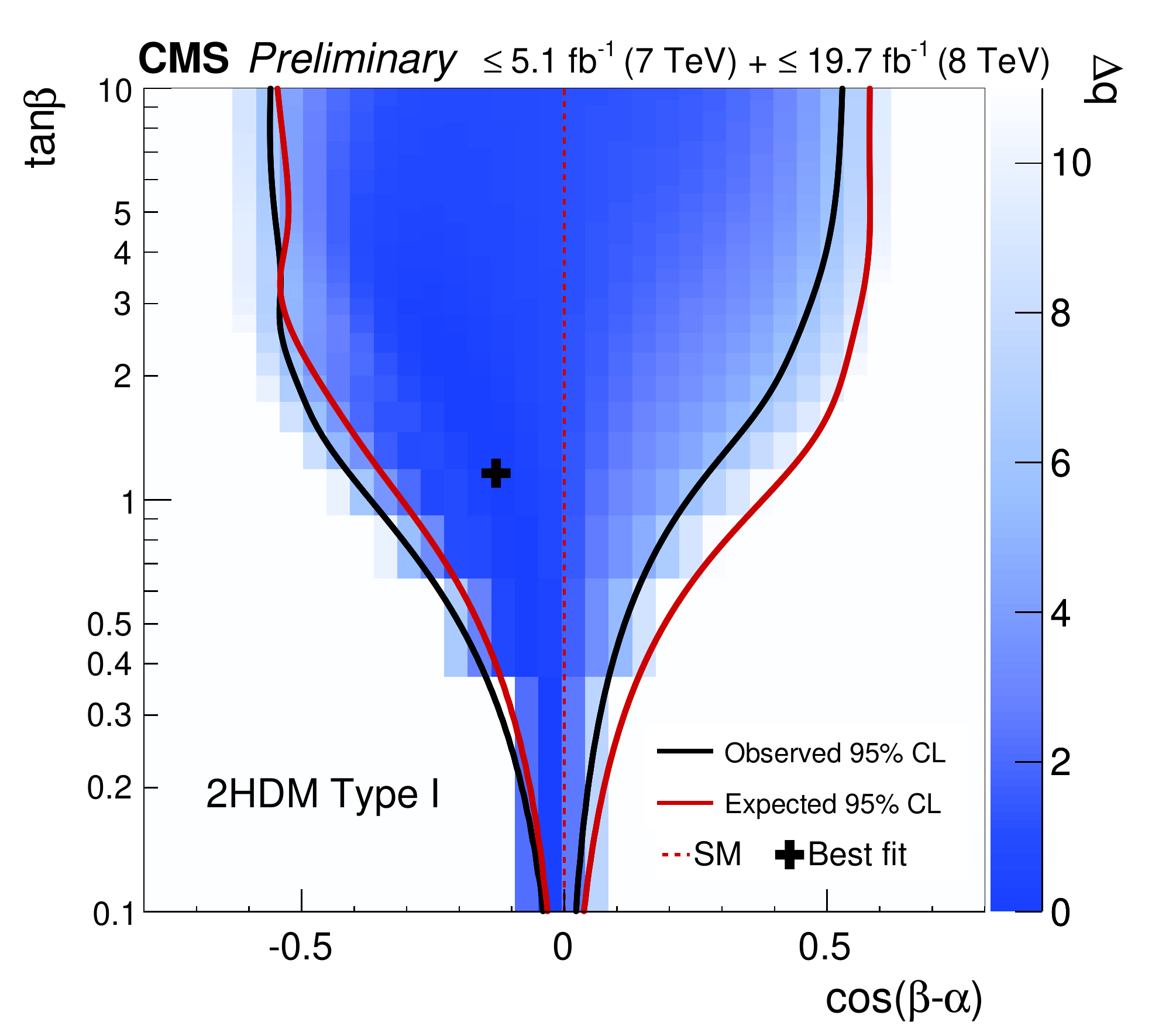}
\includegraphics[width=8.7cm, height=6.1cm]{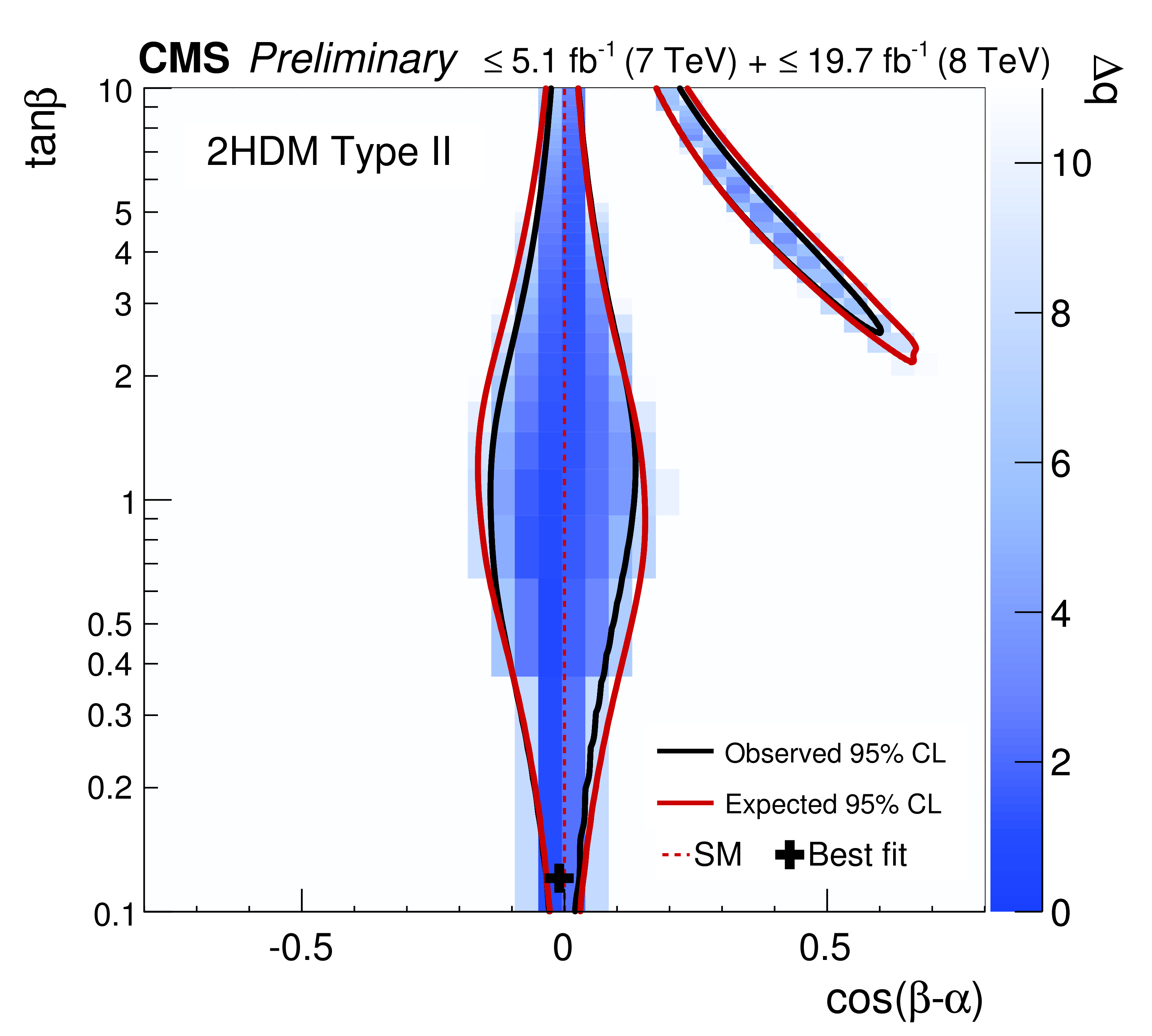}
 \caption{CMS constraints on the mixing angles $\alpha$ and $\beta$, for 2HDM-I and 2HDM-II. Figures from ref. \cite{CMS:2016qbe}.}
  \label{Fig:SMHix5}
 \end{center}
\end{figure}

 On the other hand, the Higgs coupling with the fermions ($b\bar{b}, c\bar{c}, \tau^+\tau^-$),
which could be measured at next-linear collider (NLC) with a precision 
of a few percent. This is particularly interesting for the large $\tan\beta$ region, where 
the corrections to the coupling $h\bar{b}b$ could change the
sign \cite{Modak:2016cdm}, and modify the dominant decay of the light Higgs,
as well as the associated production of the Higgs with 
b-quark pairs \cite{Balazs:1998nt, DiazCruz:1998qc}.

\section{The flavor violating Higgs and top decays signals}

In the models that we are interested in, both flavor and CP-violation could occur, either
at tree- or loop-levels. Within the 2HDM-Tx it occurs at tree-level, with large rates that make
it feasible to be searched at current and future colliders. In order to evaluate the viability of the Higgs
signals, one needs first to consider all low-energy processes, and use the current bounds to look for
allowed regions of parameter space. This is shown in figure 10, from \cite{Arroyo:2013tna}, with some
emphasis on the LFV processes. Figure 10-left shows the allowed region in the plane $\tan\beta- \cos(\beta-\alpha)$,
after LHC constraints, while figure 10-right shows the bounds from flavor-dependent processes. 
One of the most sensitive constraint is provided by the muon anomalous
magnetic moment,  which deviates from the SM, and it is difficult to reproduce with the 2HDM of type I and II,
but  for model III we do find viable regions of parameter space.  For instance, for values around
$\tan\beta\simeq 7$, the allowed heavy Higgs mass range is: $500 < m_H < 1000$ GeV; 
another region around $\tan\beta\simeq 12-15$ is also allowed, but now only for
$850 < m_H < 1000$ GeV. 

\begin{figure}[t]
\begin{center}
\includegraphics[width=7.4cm, height=6.3cm]{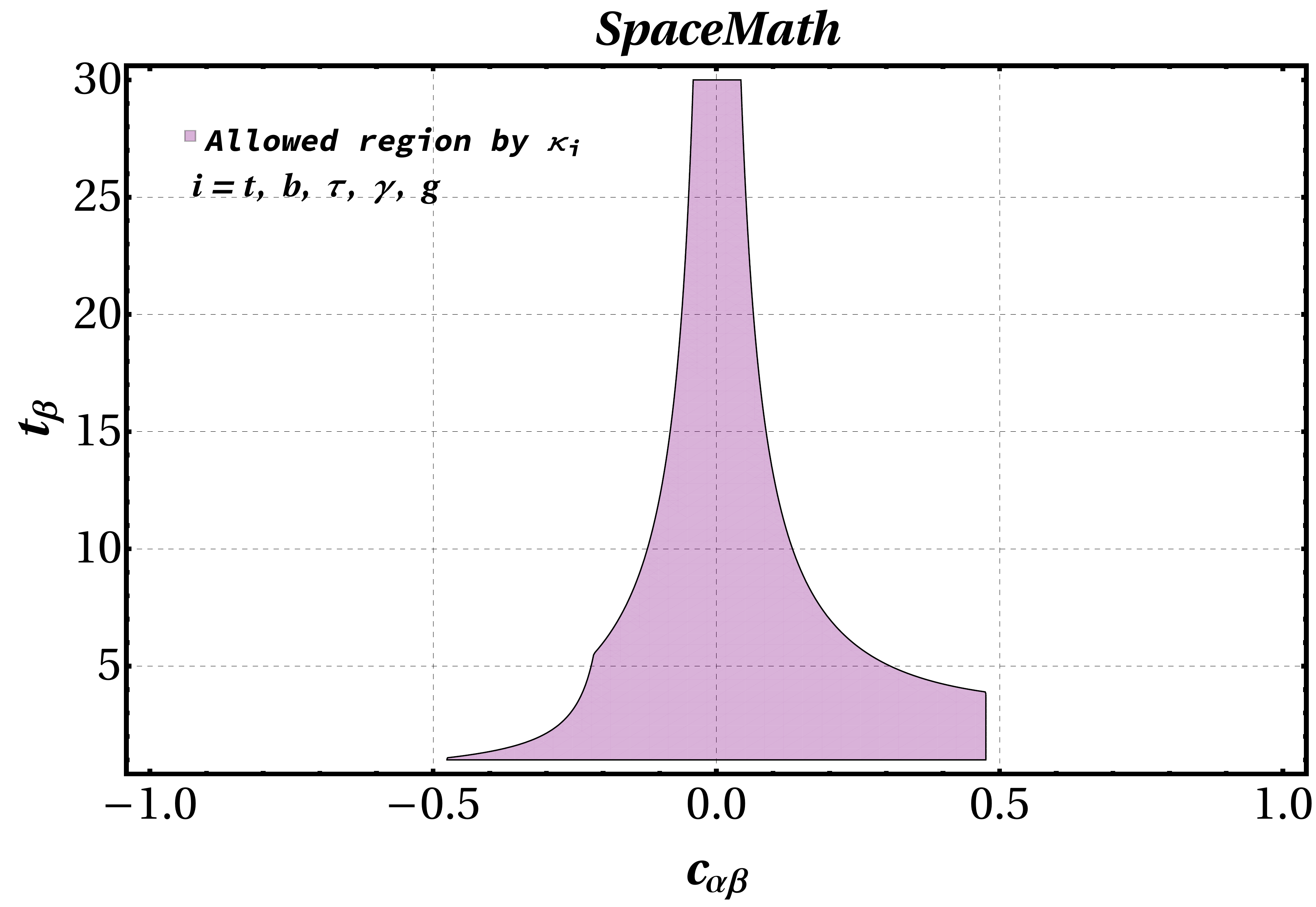}
\includegraphics[width=7.7cm, height=6.1cm]{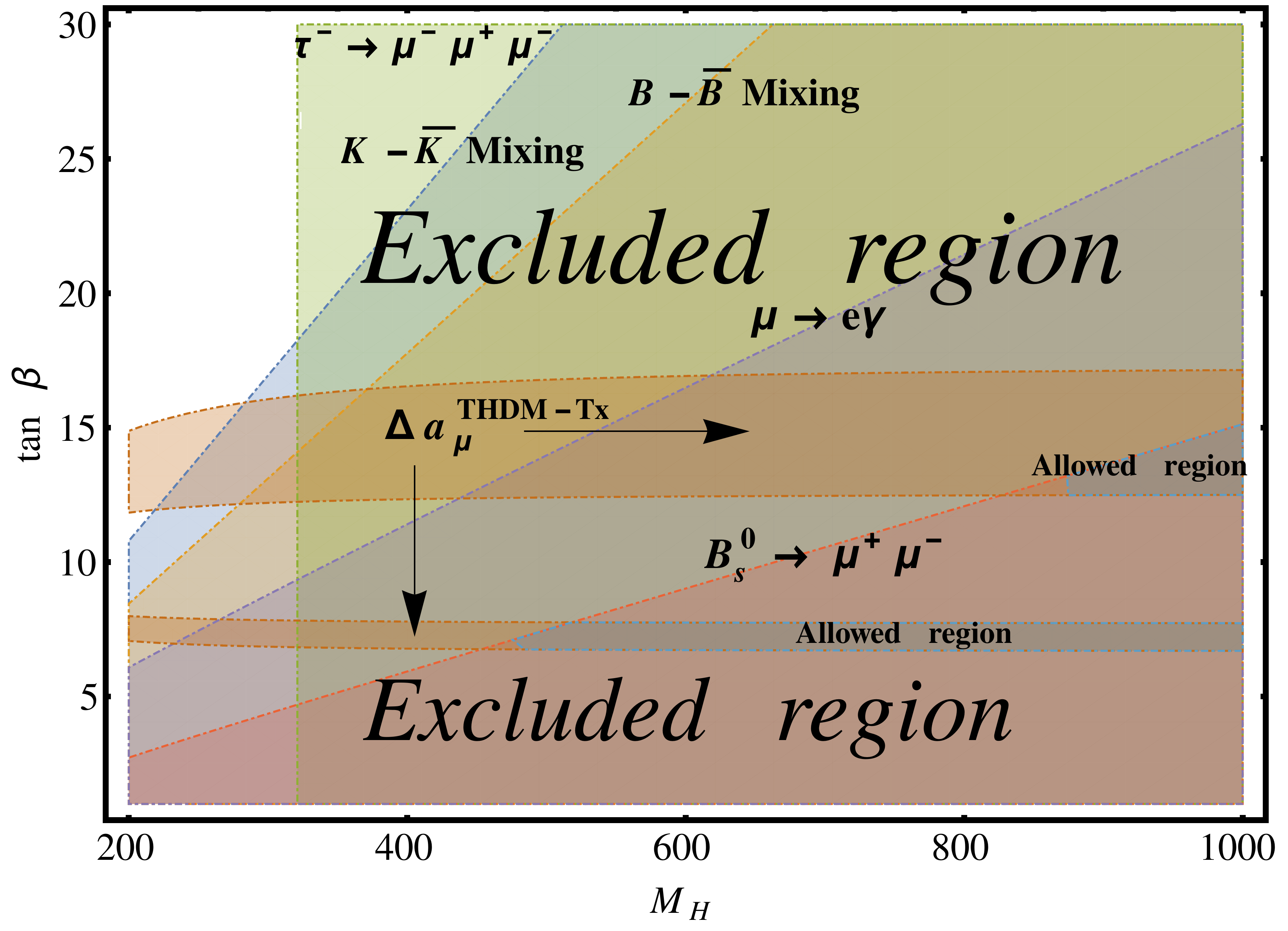}
 \caption{Constraints from flavor physics on 2HDM. Figures from \cite{Arroyo:2013tna}.}
  \label{Fig:lfvhix}
 \end{center}
\end{figure}

\subsection{LFV Higgs decays}

Within the SM, LFV processes vanish  at any order of perturbation theory, which  motivates 
the study of  SM extensions that predict sizable  LFV effects that could be at the reach of detection. 
In particular, the observation of neutrino oscillations, 
which is associated with massive neutrinos, motivates the occurrence of Lepton Flavor Violation (LFV)
in nature \cite{Valle:2018pgs,Ma:2006fn}, which could be tested with the charged lepton decays, 
 $l_i\to l_j \gamma$ and $l_i\to l_j \bar l_k l_k$.
 Another interesting possibility, which became even more relevant after the Higgs discovery,
is the  decay $h \to \tau \mu$, which was studied first in Refs. \cite{Pilaftsis:1992st,DiazCruz:1999xe}, 
with subsequent analyses on the detectability of the signal appearing soon after 
\cite{Han:2000jz, Assamagan:2002kf, Kanemura:2005hr, McKeen:2012av}. This motivated a plethora of calculations 
in the framework of several SM extensions, such as theories with massive neutrinos, supersymmetric 
theories, etc.  \cite{Cotti:2002zq, Arganda:2004bz,DiazCruz:2002er,Brignole:2004ah,DiazCruz:2008ry, Curtin:2013fra}.
(For more on LFV Higgs decays see also  \cite{Arana-Catania:2013xma, Arhrib:2012ax, Harnik:2012pb, Goudelis:2011un, GomezBock:2005hc}).  

Nowadays, the decay $h\to\tau\mu$ is included in the Higgs boson studies performed at LHC, which 
offers a great opportunity to search for  new physics at the LHC. Along this line, a slight excess of   
$h \to \tau \mu$ signal was reported at the  LHC run I, with a significance of 2.4 standard deviations \cite{Khachatryan:2015kon}, 
but subsequent studies \cite{Aad:2016blu, Sirunyan:2017xzt}  ruled out such an  excess and instead put the limit 
$BR(h\to \bar{\mu}\tau) < 1.2 \times 10^{-2}$  with 95\% C.L. 
Current bounds on LFV Higgs decays from CMS are shown in figure 11, for the plane of LFV Higgs couplings
$|Y_{\tau\mu}| -|Y_{\mu\tau}|$, which also shows the constraints obtained from LFV lepton decays;
we can see that the LFV Higgs decay provides the strongest constraints on these parameters.

\begin{figure}[t]
\begin{center}
\includegraphics[width=13.7cm, height=6.3cm]{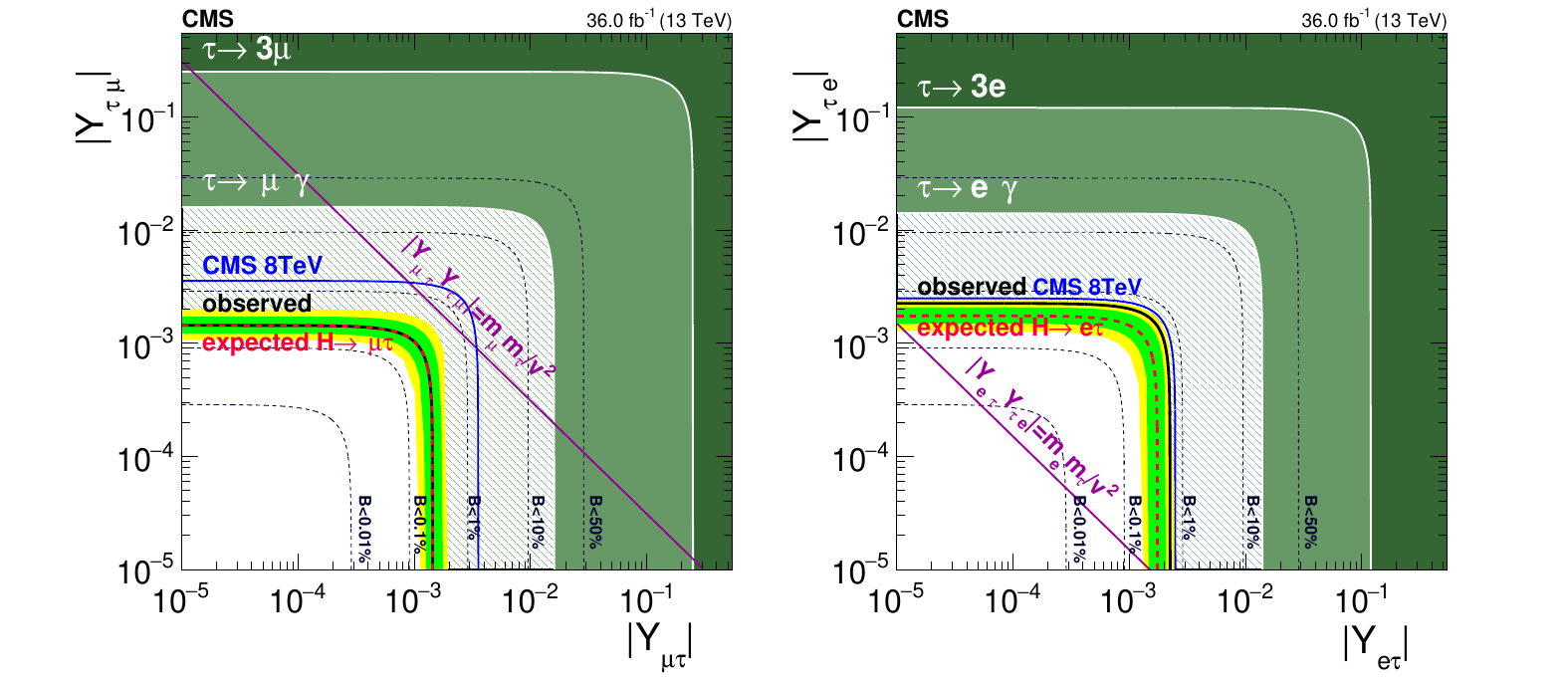}
 \caption{Bounds on LFV Higgs couplings. Figure from \cite{Sirunyan:2017xzt}.}
  \label{Fig:SMHix6}
 \end{center}
\end{figure}

Now we can use those strongest constraints on  the LFV Higgs decay, in order to test the 2HDM-Tx, for the choices
of textures that me mentioned before: Parallel, Complementary and semi-Parallel. This is shown in figure 12, 
for the cases studied in ref. \cite{Arroyo:2013tna}, and we can see that all of them
satisfy the current LHC limits, but the predicted rates could tested at the coming LHC runs..

\begin{figure}[t]
\begin{center}
\includegraphics[width=9.7cm, height=7.3cm]{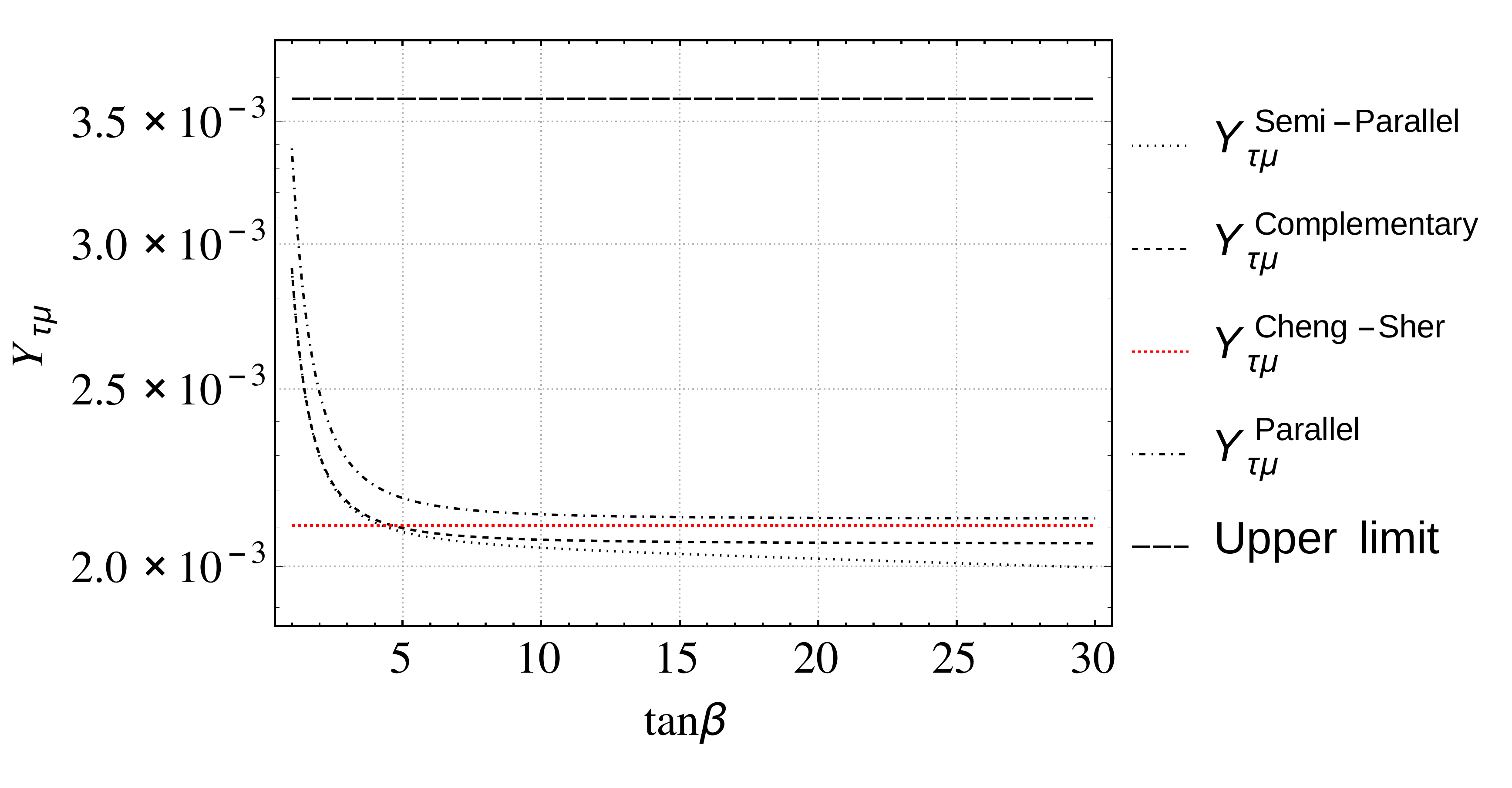}
 \caption{Constraints  on LFV Higgs couplings. Figure from \cite{Arroyo:2013tna}.}
  \label{Fig:SMHix2}
 \end{center}
\end{figure}

\subsection{FCNC top  decays}

Top quark rare decays has been studied for several years as a channel to search of 
new physics \cite{LDCtopfv, Eilam:1991, Mele:1998, Aguilar:2003, Jenkins:1996zd, Lorenzo:1999}, which
included a variety of theoretical calculation for $BR(t\rightarrow ch)$ \cite{Bar-Shalom:1998, DiazCruz:2012xa,DiazCruz:2001gf}. 
We use the following expression for the branching ratio: $BR(t\rightarrow ch)=\frac{\Gamma(t\rightarrow ch)}{\Gamma_{tot}}$,
where the total top width is given by: $\Gamma_{tot}\approx 1.55 GeV$, and the width for the FCNC
top decay (in the CPC case) is:
\begin{equation}
 \Gamma(t\rightarrow ch)\approx \frac{g^2_{htc}}{32\pi}\frac{v^2}{f^2}m_t\left(1-\frac{m^2_h}{m_t^2}\right)^2
\end{equation}
The resulting rates for $BR(t\to ch)$,  are shown in figure 13, for
the cases considered in ref. \cite{Arroyo:2013tna}.

So far, LHC has provided the limit $B.R.(t\rightarrow ch) < 2\times 10^{-3}$ \cite{Khachatryan:2014jya}. On the other hand,  
Ref. \cite{AguilarSaavedra:2000aj,topfcnc,Chen:2013qta}, provides some estimates for the branching ratios for $t\rightarrow ch$ that could be proved at the different phases of LHC. For instance, it is claimed there that top decay processes provide the best channel to discover top FCNC interactions, while only in some cases it is surpassed by single top production, when up and charm quark  FCNC interactions are involved. In some of the examples discussed in ref. \cite{topfcnc}, the maximum rates predicted to be observable, with a 3 $\sigma$ statistical significance or more,  is about is $BR < 5.8\times 10^{-5}$,  for one LHC year with a luminosity of 6000 $fb^{-1}$. This implies that the top FCNC branching ratio that arise within the 2HDM-Tx can be proved at LHC.

\begin{figure}[t]
\begin{center}
\includegraphics[width=9.7cm, height=7.3cm]{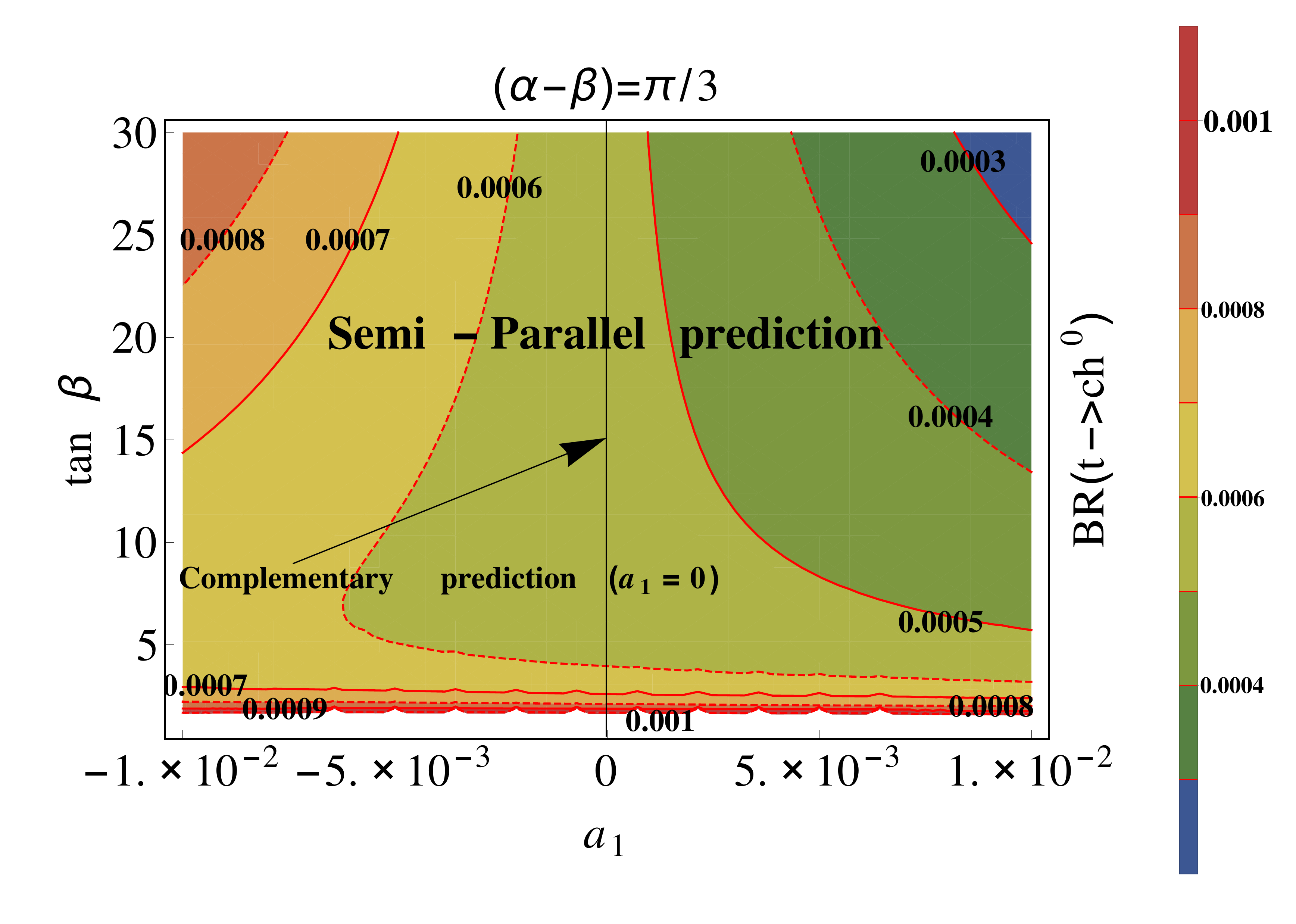}
 \caption{Constraints  on  Higgs couplings and implications $BR(t\to ch)$. Figure from \cite{Arroyo:2013tna}.}
  \label{Fig:SMHix2}
 \end{center}
\end{figure}

\section{Conclusion and outlook}

One of the most important task of future colliders is to study the properties 
of the  Higgs-like particle with $m_h=126$ GeV  discovered at the LHC.
Current measurements of its spin, parity, and interactions, seems consistent with 
the SM.
We have reviewed the essentials of the SM Higgs sector, going from the model definition up to its phenomenology.
Then, we presented the motivation for the multi-Higgs models,   with particular emphasis on the
general 2HDM. Constraints on the Higgs-fermion couplings, derived from Higgs search at LHC, and their implications
 were discussed too. In the down-quark sector, there are a
interesting aspects to study, such as the rates for rare b-decays. 
Furthermore, as a consequence of Lepton Flavor Violating (LFV) 
Higgs interactions, the decay $h \to \tau\mu$ can be induced at rates 
that could be detected at future colliders. The complementarity of future colliders has 
been studied in \cite{Kanemura:2014dea}.
For the up sector, perhaps the most interesting signal is provided by the
FCNC top decay $t\to ch$, which can be studied at LHC.

Within the SM we have only one Higgs doublet giving masses to all type of fermions,
and thus its couplings to fermions and gauge bosons are proportional to the 
particle mass, and they lay on a single line, when plotted as function of the particle mass. 
But it is desirable to test this property and study models where more than one Higgs multiplet
participates in the fermion mass generation. In fact, the THDM is one example where such
scheme arises; in this case we have that the fermion couplings lay on two lines, one for up-type 
quarks and one for leptons and down-type quarks. There are also models where
each fermion type gets its mass from a private Higgs, for this we need at least 3 Higgs doublets.
In this case, the Higgs couplings as function of the fermion mass lays on three different lines.
We have also studied these extended Higgs sector,  both for a Non-SUSY model \cite{Diaz-Cruz:2014pla} 
and for a SUSY model (4HDM) \cite{Diaz-Cruz:2019emo}. 
Besides serving us as an specific model to test the pattern of Higgs couplings,
and new physics, this model can also be motivated from the scenarios having one Higgs for 
each generation, such as the $E_6$ GUT model and the superstring-inspired models \cite{Aranda:2000zf}. 

 To test this type of  models one needs
to measure Higgs couplings with fermions of at least two generations. So far, LHC has measured
couplings with fermions of the 3rd generation, and there is some chance to measure couplings
with charm and muons, which will permit to test the private Higgs hypothesis.
At linear colliders it will be possible to measure these couplings with better precision; 
more recently there have appeared some claims that the coupling with strange quarks is also
possible, both the flavor-conserving \cite{Duarte-Campderros:2018ouv} and flavor-violating ones \cite{Barducci:2017ioq}.
 Furthermore, this could also be possible within the models with Higgs portal to dark matter,
In such case, the direct detection of DM depends on the  Higgs interaction with nucleons,  
which in turn depends on the Higgs coupling with light quarks. 
Therefore, by searching for DM-nucleon dispersion one is testing the Higgs coupling
with light quarks. Similar remarks hold when one considers $e-\mu$ conversion, where the
Higgs nucleon interaction also plays a role. 

Thus,  nature could be extra benevolent, and  by permitting the existence of a light Higgs boson,
it could have provided us with a tool to search for physics beyond the SM. The detection of the Higgs boson could
be of such relevance, that it may be the key to find what lies beyond the SM, with implications ranging from flavor physics 
to dark matter, supersymmetry and even cosmology \cite{Espinosa:2013lma}.

\bigskip

\noindent \underline{\it Acknowledgements}

I acknowledge support from CONACYT-SNI (Mexico) and VIEP(BUAP). I would like to thank M.A. Perez, G.L. Kane, and
A. Mendez, for their support and many lessons on Higgs physics. Special thanks to my classmate C.P. Yuan, and to
 A. Rosado, G. Lopez-Castro and A. Aranda, for many years of fruitful collaborations. I also learned from my former 
postdocs and students  on the topics that appear in this paper: M. Arroyo, J. Halim, B. Larios,  J. Hernandez, Olga Felix, R. Noriega-Papaqui,
J. Orduz, U. Saldana. M.Arroyo and B. Larios helped me with some of the figures, and the higgsitos M. Prez de Leon and R. Garcia
assisted me by reading the draft and suggested some corrections.

\baselineskip 16pt
\bibliographystyle{unsrt}

\end{document}